\documentclass[twocolumn,aps,superscriptaddress,pre]{revtex4}

\usepackage{amsmath,amssymb,graphicx}
\usepackage{subfigure}
\usepackage{algorithmic,algorithm}
\usepackage{enumerate}
\usepackage{times}
\usepackage{color}
\usepackage{soul}
\allowdisplaybreaks[4]
\definecolor{yblue}{rgb}{0.06, 0.3, 0.57}
\usepackage[pdftex]{hyperref}
\hypersetup{colorlinks=true,linkcolor=blue,citecolor=blue,urlcolor=blue}

\newcommand{\state}[1]{|{#1}\rangle}
\newcommand{\statem}[1]{$|{#1}\rangle$}

\newcommand{\statelp}[1]{$|{#1}\rangle_P$}
\newcommand{\statec}[1]{$\mathcal{C}_{#1}$}
\newcommand{\statep}[1]{$\mathcal{P}_{#1}$}

\begin{document}

\title{Linear limit continuation: Theory and an application to two-dimensional Bose-Einstein condensates}

\author{Wenlong Wang}
\email{wenlongcmp@scu.edu.cn}
\affiliation{College of Physics, Sichuan University, Chengdu 610065, China}


\begin{abstract}
We present a coherent and effective theoretical framework to systematically construct numerically exact nonlinear solitary waves from their respective linear limits. First, all possible linear degenerate sets are classified for a harmonic potential using lattice planes. For a generic linear degenerate set, distinct wave patterns are identified in the near-linear regime using a random searching algorithm by suitably mixing the linear degenerate states, followed by a numerical continuation in the chemical potential extending the waves into the Thomas-Fermi regime. The method is applied to the two-dimensional, one-component Bose-Einstein condensates, yielding a spectacular set of waveforms. Our method opens a remarkably large program, and many more solitary waves are expected. Finally, the method can be readily generalized to three dimensions, and also multi-component condensates, providing a highly powerful technique for investigating solitary waves in future works.
\end{abstract}


\maketitle

\section{Introduction}
Solitary waves are ubiquitous in rather diverse physical systems ranging from normal fluids to Bose-Einstein condensates (BECs), superfluids and superconductors, and nonlinear optics \cite{becbook1,becbook2,Panos:book,Nicolay:SF,DSoptics}. Particularly, atomic BECs have provided a pristine platform for investigating solitary waves due to the powerful experimental techniques available to manipulate and control the cold atoms. Importantly, the mean-field treatment is frequently relevant as the atomic gas is ultradilute. The interactions between atoms as well as the trap potential can be tuned flexibly. Cold atoms also provide novel opportunities to study rotational condensates \cite{fetter2,RCG:rotating}, condensates in optical lattices \cite{BECOL}, and vector solitary waves in both pseudo-spinor and spinor condensates \cite{Becker:DB,3CBECexpt}. As such, a detailed exploration of solitary waves in this context, e.g., their existence, stability, dynamics, and interactions, is of paramount importance for understanding quantum fluids, investigating solitary wave pattern formations, as well as developing a broad spectrum of theoretical and computational methods \cite{PK:VXreview,Wang:AI,Panos:DC1,Panos:DC2,Panos:DC3}.

Pattern formation is a very fascinating phenomenon, which is partially responsible for the transition from elementary structures to complex systems. Two nucleons, namely protons and neutrons, can combine into hundreds of atomic nuclei. A few dozens of atoms, via the various chemical bonding processes, form a tremendous number of molecules and crystals, which is a major theme of chemistry. While solitary waves are different in nature from these ``elementary'' particles, they nonetheless exhibit a similarly rich pattern formation \cite{Panos:DC1,Panos:DC2,Panos:DC3,Brand_SV}. A novel feature of the pattern formation here is that the basic structures may be flexibly deformed.

Numerous solitary waves have been studied in atomic BECs, encompassing but not limited to bright solitons in attractive condensates \cite{tomio}, dark solitons in repulsive condensates \cite{Dimitri:DS}, gap solitons in optical lattices \cite{BECGS}, and vortical structures in both scalar and vector condensates \cite{Ionut:VR,Wang:VR,Wang:VRB}.  In this work, we focus on the common repulsive scalar condensate.
Two prototypical types of solitary wave excitations exist in this setting, i.e., the dark soliton surface and the vortical filament, due to the nature of the complex scalar field. Both excitations feature a localized density dip, with a phase jump of $\pi$ across a stationary dark soliton surface and a long-range phase winding of $2\pi$ around a vortical filament. Dark soliton surfaces and vortical filaments cannot terminate in the fluid, but they can close on themselves or terminate at the fluid boundary. Depending on the trap geometry, they can exhibit various forms and even acquire different names.
In a quasi-$2d$ pancake condensate, they can manifest as effective dark soliton filaments and point vortices, respectively \cite{Wang:AI,PK:VXreview}. In a quasi-$1d$ cigar condensate, dark solitons can also arise as effective point particles \cite{Dimitri:DS}. Complicated crossover settings are also possible, e.g., when the transverse modes of a cylindrical condensate are excited, solitonic vortex states are found \cite{Chladni,Brand_SV}. As this work focuses on numerical method, we study the quasi-$2d$ condensate herein for simplicity. Nevertheless, the pattern formation remains extremely rich and diverse, despite that we ``only'' have dark soliton filaments and point vortices in this limiting setting.


To study the pattern formation of solitary waves and their properties, there is much interest in finding numerically exact stationary states. This is particularly so considering that analytic methods are essentially limited to the $1d$ homogeneous (integrable) setting \cite{Lichen:DT}. However, we mention in passing that approximate theoretical methods based on physical insights are available, e.g., the two-mode analysis in the near-linear and intermediate regimes \cite{Panos:TMA,PK:DSVX}, and reduced particle-level dynamics in the Thomas-Fermi (TF) regime \cite{PK:VXreview,Wang:AI}.
Numerically exact solutions also offer insight into the interesting dynamics of the solitary waves via the Bogoliubov-de Gennes (BdG) linear stability analysis, and direct time evolution. Both robust oscillations and symmetry-breaking instabilities of a solitary wave can be studied in detail. 
A pioneering method to find numerically exact states at the systematic level is the deflation method \cite{Panos:DC1,Panos:DC2,Panos:DC3}. An entirely different approach is to construct solitary waves from the linear limits by a numerical continuation in the chemical potential.



The technique of constructing solitary waves from their linear limits has been intensively employed in previous works to study \textit{particular} states by physical insights \cite{Carr:VX,Wang:DSS,Wang:VR}. Recently, we have successfully developed this method into a \textit{systematic} one in the $1d$ setting \cite{Wang:DD,Wang:MDDD,Wang:DAD}. Numerous solitary waves of increasing complexity are found in $1d$ upto a total of five components \cite{Wang:DD,Wang:MDDD}. The method was also extended further to the two-component system with different dispersion coefficients, generating yet new series of solutions \cite{Wang:DAD}. The linear limit continuation appears to hold considerable promise for identifying and classifying solitary waves. For example, the well-known dark-bright, dark-dark, dark-anti-dark, dark-multi-dark waves, and their more complex generalizations in the two-component condensates are all found in this single theoretical framework \cite{Wang:DD,Wang:DAD}.


Linear limit continuation is significantly more versatile in two and three dimensions due to the emergence of degenerate states. This new feature makes the problem very different in nature from that of $1d$, where linear degenerate states do not occur for bound states. For a given linear degenerate set, one may continue each basis state. In addition, certain linear combinations of the basis states can yield fundamentally new wave patterns. A famous example is the vortex state. In the $2d$ isotropic harmonic potential, the linear Cartesian states \statem{10} and \statem{01} are degenerate, leading to dark soliton stripes. However, they are \textit{not} complete, the complex mixing $(\state{10} \pm i\state{01})/\sqrt{2}$ yield the vortex and anti-vortex states, respectively. It is therefore an outstanding question how to effectively find the good linear combinations. While the theoretical Lyapunov–Schmidt reduction technique is developed for this purpose, it is frequently rather tedious to apply even to low-lying states \cite{RCG:rotating}.

The main purpose of this work is to examine the ``degenerate state problem'' and present a coherent and effective semi-analytical framework to systematically identify and continue solitary waves from their linear limits. The motivation and the details are clearly discussed, and then the method is illustrated and applied to $2d$ and a diverse set of solitary waves including many previously undiscovered ones are found in an organized manor. 
The first step is to classify the different linear degenerate sets. This is straightforward and can even be visualized by ``lattice planes'' of the quantum numbers of the Cartesian states. While this idea is very simple, it is highly significant in our opinion as it makes the classification particularly transparent. One can then study each linear degenerate set in a one-by-one manner, in line with the idea of divide and conquer. The second central step is to construct distinct solitary waves from a given set of degenerate states. A key numerical observation is that the number of distinct solitary waves bifurcating from a finite set of linear degenerate states is finite, which motivates us to propose a random solver. A suitable random linear combination serves as an initial guess of the Newton's solver in the near-linear regime, and the process is repeated for a number of independent runs. Next, we identify distinct states, and they are subsequently continued further in the chemical potential into the TF regime. The details of the method should be discussed in the next section. An additional feature in $2d$ is the presence of chemical potential chaos, which is not frequently encountered in $1d$. This refers to that the equilibrium configuration of a solitary wave may morph nontrivially in the chemical potential evolution. As such, we also summarize and illustrate the typical chaotic processes in this work.

We highlight that the ``degenerate state problem'' is the central and crucial step for developing the linear limit continuation method. Once this is solved, the entire theoretical framework is essentially complete. Indeed, the generalization to $3d$ is straightforward at least conceptually. In addition, the techniques developed for multi-component systems in $1d$ can also be readily incorporated into $2d$ and $3d$ systems. We shall discuss these later in Sec.~\ref{co}.

This work is organized as follows. We present the theoretical and numerical setup in Sec.~\ref{setup}. The method is applied to the $2d$ scalar condensate, and the low-lying solitary waves bifurcating from two prototypical lattice planes are illustrated and discussed in Sec.~\ref{results}. Finally, a summary and an outlook of future directions are given in Sec.~\ref{co}.




\section{Theoretical and numerical setup}
\label{setup}

\subsection{Simple linear limit continuation}
We start from the following dimensionless Gross-Pitaevskii equation in two dimensions:
\begin{align}
    -\frac{1}{2} \nabla^2 \psi+V \psi +| \psi |^2 \psi = i \frac{\partial \psi}{\partial t},
    \label{GPE}
\end{align}
where $\psi(x,y,t)$ is the macroscopic wavefunction, $V=(\omega_x^2 x^2 + \omega_y^2y^2)/2$ is the harmonic potential. It is convenient to set $\omega_y=1$ by scaling without loss of generality, and $\kappa=\omega_y/\omega_x$ is the trap aspect ratio. For the radially symmetric potential, we further define $\omega=\omega_x=\omega_y=1$. Stationary state of the form $\psi(x,y,t)=\psi^0(x,y) e^{-i\mu t}$ leads to:
\begin{align}
    -\frac{1}{2} \nabla^2 \psi^0 +V \psi^0 +| \psi^0 |^2 \psi^0 = \mu \psi^0,
    \label{GPEstationary}
\end{align}
where $\mu$ is the chemical potential. This is the central equation we aim to solve in this work. This equation has a few generic symmetries. First, it has the $U(1)$ symmetry, if $\psi^0$ is a solution, then $\exp(i\theta_0)\psi^0$ is also a solution for any real global phase shift $\theta_0$. Second, it has the charge conjugation symmetry, $\psi^{0*}$ is also a solution. Third, it bears the reflection symmetry with respect to both the $x$ and $y$ axes, i.e., a reflection of $\psi^0$ over either axis remains a valid solution. Finally, it bears the full rotational symmetry in the isotropic trap with respect to the trap center, i.e., a rotated $\psi^0$ about the origin is also a solution. Here, we find distinct stationary waves upto these underlying symmetries.

The underlying linear limit is the two-dimensional quantum harmonic oscillator which can be readily solved analytically in the Cartesian coordinates. In the isotropic case, solutions in the polar coordinates are also available. In principle, the polar basis is not strictly necessary, one basis is sufficient. We favour the Cartesian basis because it also works for anisotropic traps. However, we also frequently refer to the polar basis to gain more physical insights and for convenience, because a linear state can be complicated in one basis but become simple in the other basis, see also below for numerical method considerations. The respective normalized linear states in Cartesian coordinates and polar coordinates read:
\begin{align}
&\varphi_{n_x,n_y}^0(x,y) = \frac{H_{n_x}(\sqrt{\omega_x}x)H_{n_y}(\sqrt{\omega_y}y)}{\sqrt{\pi2^{n_x}n_x!2^{n_y}n_y!}\kappa^{1/4}} e^{-(\omega_x x^2 + \omega_y y^2)/2}, \label{linearsc} \\
&\varphi_{n_r,n_{\theta}}^0(r,\theta) = \sqrt{\frac{n_r!}{\pi (n_r+|n_{\theta}|)!}} L_{n_r}^{|n_{\theta}|}(r^2) r^{|n_{\theta}|} e^{-\omega r^2/2}e^{in_{\theta} \theta}. \label{linearsp}
\end{align}
Here, $H$ is the Hermite polynomial and $L$ is the associated Laguerre polynomial. The corresponding eigenenergies are $E_{n_x,n_y}=(n_x+1/2)\omega_x+(n_y+1/2)\omega_y$, where $n_x, n_y = 0, 1, 2, ...$ and $E_{n_r,n_{\theta}}=(2n_r+|n_{\theta}|+1)\omega$, where $n_r, |n_{\theta}|=0, 1, 2, ...$ The four quantum numbers in turn represent linear dark nodes or cuts along $x$, linear cuts along $y$, radial circular cuts, and the central vorticity. The $|n_\theta|$ also represents equally spaced angular cuts passing through the origin in the sense of both the real and the imaginary parts. For convenience, $\state{mn}_C$ or $\state{mn}$ represents a linear state in the Cartesian basis, and $\state{rt}_P$ stands for a linear state in the polar basis. Their corresponding solitary waves are then denoted as \statec{mn} and \statep{rt} if relevant, respectively.

The polar states are also important for benchmarking numerical methods. They are rotationally symmetric upto a topological charge, therefore, they can be efficiently solved in the reduced $1d$ setting. Assuming that a nonlinear state \statep{nm} takes the form $\psi^0=u_n^0(r)\exp(im\theta)$, then Eq.~\eqref{GPEstationary} becomes:
\begin{align}
    -\frac{1}{2} (\frac{d^2}{dr^2}+\frac{1}{r}\frac{d}{dr}-\frac{m^2}{r^2}) u_n^0 +V(r) u_n^0 +| u_n^0 |^2 u_n^0 = \mu u_n^0.
    \label{GPE1d}
\end{align}
The $1d$ solutions can be typically computed to much higher accuracy, serving as benchmarking solutions for the $2d$ calculation. As such, the $2d$ numerical setup and simulation parameters should be sufficiently good to reproduce the pertinent polar states in a linear degenerate set before applied to the other states. Finally, a $2d$ state can be obtained from the $1d$ solution using, e.g., a cubic spline interpolation. We mention in passing that the BdG spectrum can also be computed in the reduced framework using the partial-wave method \cite{Carr:VX,Wang:DSS}, and similar reductions to $2d$ and even $1d$ are available in $3d$ depending on the symmetry of the pertinent states \cite{Wang:RDS,Wang:DSS}.

\begin{figure*}[t]
\includegraphics[width=\textwidth]{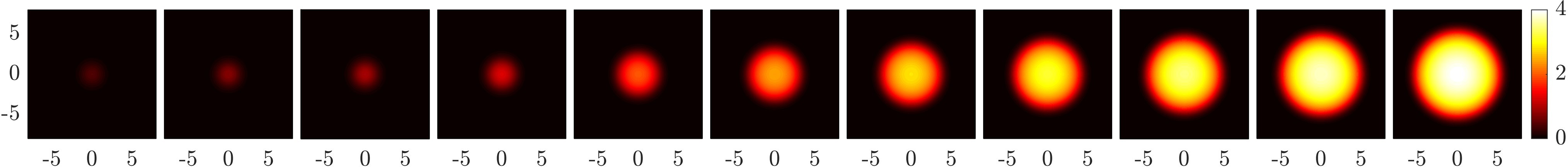}
\includegraphics[width=\textwidth]{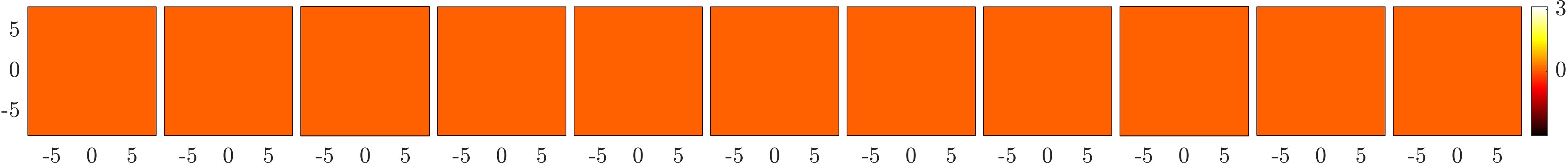}
\includegraphics[width=\textwidth]{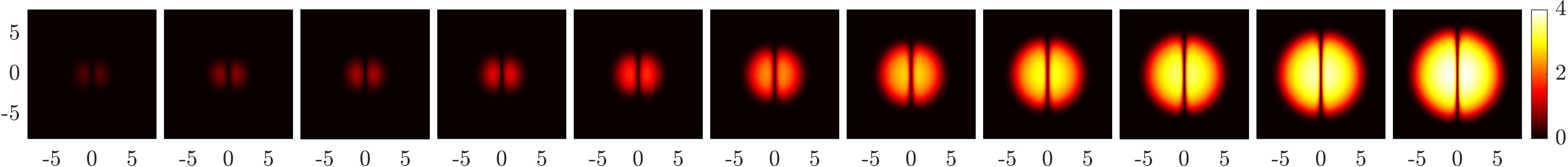}
\includegraphics[width=\textwidth]{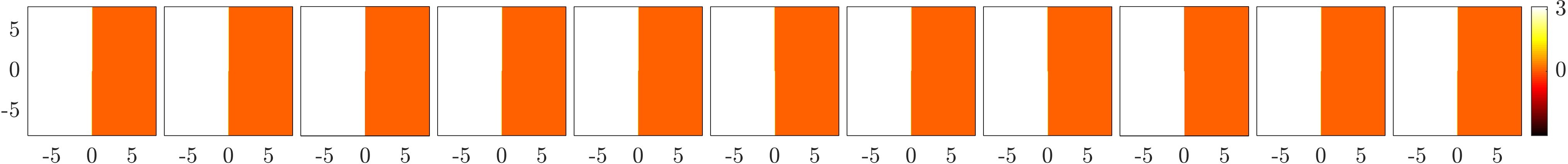}
\includegraphics[width=\textwidth]{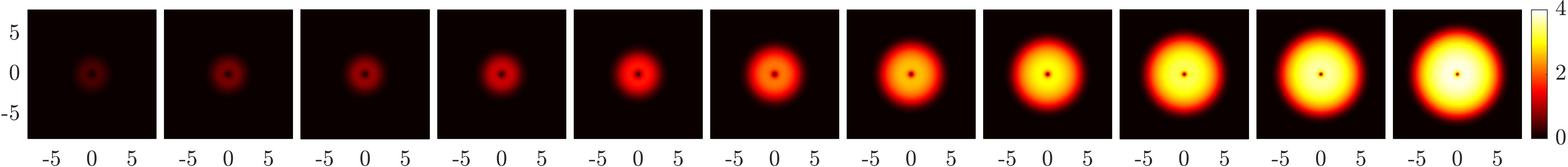}
\includegraphics[width=\textwidth]{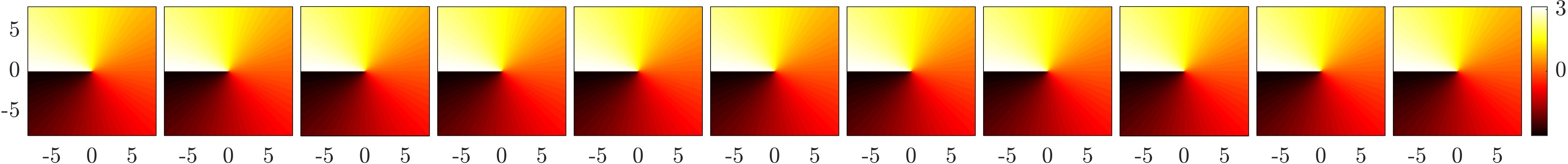}
\caption{
Illustration of the linear limit continuation in the chemical potential in turn for the ground state, the dark soliton stripe state, and the vortex state, the odd and even panels depict the magnitude and phase profiles, respectively. The ground state has a linear limit from \statem{00} at $\mu_0=1$, the states at chemical potentials $\mu=1.1, 1.3, 1.5, 2, 4, 6, 8, 10, 12, 14, 16$ are depicted, showing the continuation from the low-density near-linear regime to the high-density TF regime. The dark soliton and vortex states have their respective linear limits from \statem{10} and \statelp{01}$=(\state{10}+i\state{01})/\sqrt{2}$ both at $\mu_0=2$, the states at $\mu=2.1, 2.3, 2.5, 3, 4, 6, 8, 10, 12, 14, 16$ are illustrated. In the TF regime, a dark soliton stripe and a vortex are respectively embedded in the TF ground state. 
}
\label{GS}
\end{figure*}

The linear limit continuation in the near-linear regime can be appreciated in the framework of the perturbation theory. For simplicity, we now consider an isolated or nondegenerate linear stationary state $\varphi^0$ with eigenenergy $\mu_0$, and we shall discuss the degenerate states later. Perturbation analysis suggests the following approximate nonlinear stationary solution to leading order in the entire near-linear regime:
\begin{align}
    \psi^0 &=\sqrt{\epsilon}\varphi^0, \\
    \mu &= \mu_0 + \epsilon \mu_r, \\
    \mu_r &= \iint |\varphi^0|^4 dx dy,
\end{align}
where $\epsilon$ is a small perturbation parameter. The idea is to use the above perturbation ansätze to self-consistently estimate the quantity $\mu_r = \iint |\varphi^0|^4 dx dy$ by averaging over the space. For a given $\mu \gtrsim \mu_0$, one finds a good representation of the nonlinear wave. On the numerical side, it provides a good initial guess for finding a numerically exact solution near the linear limit. The converged solution is subsequently used as the initial guess for yet a slightly larger chemical potential. The process repeats following a prescribed chemical potential schedule and a series of numerically exact solutions is obtained if successful, i.e., the wave is continued from the near-linear regime into the TF regime. In practice, one may apply the perturbation initial guess a few times if necessary before the continuation runs on its own, because the wave density typically grows rapidly in the vicinity of the linear limit.

The chemical potential continuation is illustrated for three well-known low-lying states, the ground state, the rectilinear dark soliton stripe, and the single vortex of charge $1$ in Fig.~\ref{GS}. The linear limit continuation nature is evident, as each state starts from a very faint field in the vicinity of the linear limit, and then gradually grows larger as the chemical potential increases. The healing length of the dark soliton and the vortex also gradually decreases with increasing density. In the TF regime, the ground state serves as a background in which the solitary waves are embedded. 

Finally, we summarize the basic technical details of the numerical methods for finding stationary states and the numerical continuation, the theoretical framework of the continuation from linear degenerate states is discussed in Sec.~\ref{CLDS}. We solve a stationary state using the finite element method for a spatial discretization and the Newton's method for convergence. We adopt a $3\times 3$ nine-point Laplacian with a square grid, and a typical spatial step size is $\delta x=0.04$. We mention in passing that we find the nearest-neighbour five-point Laplacian is \textit{not} sufficient for more excited states such as the two ring dark solitons (RDSs) in the TF regime. The domain is chosen to be sufficiently large for the condensate and then the zero boundary condition is applied. A typical spatial domain is $[-8, 8] \times [-8, 8]$. We use a piecewise constant $\delta \mu$ continuation schedule for simplicity, a typical $\delta \mu \sim O(0.01)$ or smaller. Different solitary waves may require different $\delta \mu$ values, and a single solitary wave may also require a finer $\delta \mu$ in certain chemical potential intervals when continuation chaos occurs, see Sec.~\ref{cc}. 
If a continuation step fails and requires a finer step, we restart the continuation from a successfully converged state with a smaller $\delta \mu$. An adaptive schedule based on a distance metric of the wavefunction and the chemical potential was also considered \cite{Carr:VX}. 
To benchmark the simulation parameters, we have carefully continued the pertinent polar states in both the $1d$ and the $2d$ frameworks, requiring that the results are consistent within the numerical accuracy.


\subsection{Continuation from linear degenerate states}
\label{CLDS}
The key new ingredient in $2d$ and $3d$ is the emergence of linear degenerate states, and their suitable linear combination or mixing may offer fundamentally novel states from the basis states. Therefore, it is \textit{not} sufficient to choose a linear basis and continue each basis state one by one. The linear limit continuation is typically quite robust in $1d$, where the eigenstates are nondegenerate. In this setting, one can simply continue each linear eigenstate one by one. As mentioned earlier, this approach has also been partially applied in $2d$ and $3d$ to particular states, including even quite complicated ones based on physical insights. In fact, the single dark soliton stripe and the single vortex states above are two such examples. To explore the full efficiency of the method clearly requires a systematic approach. First, there is no guarantee in general that a linear waveform by physical insights is a good initial guess for a nonlinear wave. Second, this approach does not address the completeness of finding distinct solitary waves bifurcating from a linear degenerate set. For example, the mixing $(\state{20}+i\state{02})/\sqrt{2}$ yields a square vortex quadruple, it differs from both the Cartesian and the polar basis states. In addition, this waveform is actually an \textit{approximation} as we shall discuss later, see Eq.~\eqref{SVQ} for the numerically exact state. While the linear states by physical insights can be pretty good or even exact, they may also be far from perfect, and sometimes even a subtle difference can matter \cite{Wang:RDS}. Next, we classify the possible linear degenerate states, and then discuss how to find the good linear combinations of a generic linear degenerate set.


The linear degenerate states can be classified systematically and also represented geometrically using the ``lattice planes'' of crystallography, because of the linear dispersion of the harmonic oscillator $E_{n_x,n_y}=(n_x+1/2)\omega_x+(n_y+1/2)\omega_y$. 
The quantum numbers $\{(n_x, n_y)\}$ form a quarter ``lattice''. Then the linear degenerate sets can be represented as parallel lattice planes, or in fact lines in the $2d$ case, similar to the crystal lattice planes of solid state physics. A family of lattice planes $[p, q]$ corresponds to a trap aspect ratio $\kappa=\omega_y/\omega_x=q/p$, which is a rational number and we can set $\kappa \leq 1$ without loss of generality. The lattice planes trace all the possible sets of degenerate states. It should be noted that this is not a genuine ``lattice'' because of the restriction $n_x, n_y \geq 0$, otherwise, any degenerate set should have an infinite number of states. Consequently, a typical lattice plane has a finite number of states or degeneracy, and the degeneracy tends to grow with increasing eigenenergy $\mu_0$. The ground state is clearly always nondegenerate. Interestingly, it can happen that a particular lattice plane in its family may strike through no lattice point in the restricted region, in this case, this nonphysical degenerate set is naturally skipped. In spite of this minor defect, we adopt the pertinent terminologies here for convenience, as no confusion can actually arise.

\begin{figure}
\includegraphics[width=0.49\columnwidth]{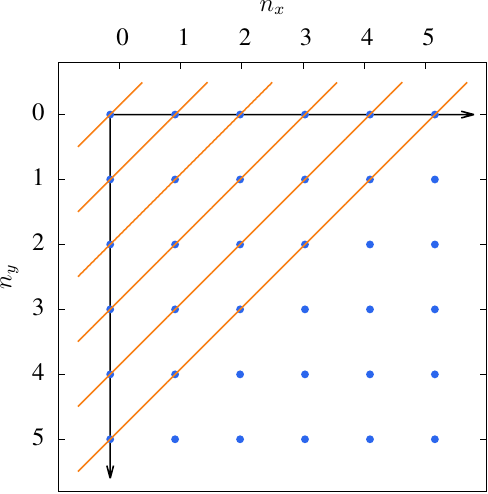}
\includegraphics[width=0.49\columnwidth]{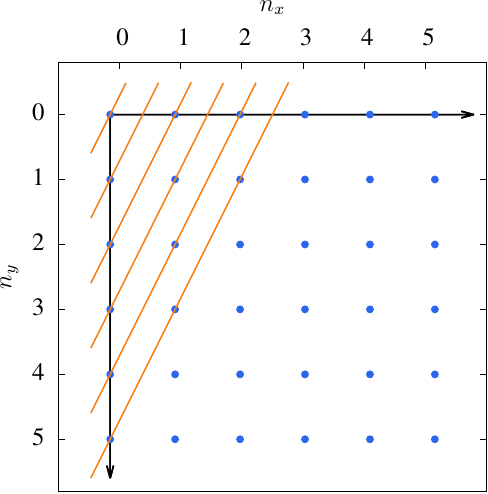}
\caption{Linear degenerate states can be geometrically represented by lattice planes, focusing only on the $n_x, n_y \geq 0$ region. Here, the low-lying linear degenerate sets of two prototypical families of lattice planes $[1, 1]$ (left panel) and $[2, 1]$ (right panel) are illustrated, and the two scenarios correspond to the trap aspect ratios $\kappa=1$ and $1/2$, respectively. In general, the lattice plane $[p, q]$ corresponds to the trap aspect ratio $\kappa=q/p \leq 1$.
}
\label{XY}
\end{figure}

The low-lying linear degenerate sets of two prototypical families of lattice planes are illustrated in Fig.~\ref{XY}. The lattice plane $[1,1]$ corresponds to the isotropic trap, the ground state has a degeneracy $1$, the first excited states have a degeneracy $2$ including \statem{10} and \statem{01}, the second excited states have a degeneracy $3$ including \statem{20}, \statem{11} and \statem{02}, and so on. The lattice plane $[2,1]$ corresponds to the trap aspect ratio $\kappa=1/2$, the ground state is again unique, the first excited state is also unique \statem{01}, the second excited states have a degeneracy $2$ including the states \statem{10} and \statem{02}, and so on. 
It is worth mentioning that while these states are continued from their linear limits with different $\kappa$ values, they are not necessarily limited to their specific $\kappa$ values. Once a state departs from its linear limit, one can further continue the $\kappa$ itself as it is also a parameter of Eq.~\eqref{GPE}, see Sec.~\ref{SAT}. Indeed, one can typically tune $\kappa$ in a wide range in the TF regime, and a continuation to other traps is also possible \cite{Wang:VRB,Wang:VB}.

The central question now is how to construct distinct solitary waves from a generic linear degenerate set. Earlier works suggest that each basis state \statem{mn} in a degenerate set can be continued solely on its own either exactly or approximately in most cases. 
While our systematic method does \textit{not} depend on this heuristic approach, it can nevertheless provide useful physical insights for constructing a large number of specific solitary waves, as patterns may exist between these states. In addition, the converged states should be found by any coherent systematic method, providing a test criterion. To this end, we present our first remark on the linear limit continuation.



\textbf{Remark 1.} It is helpful to continue from different possible basis states if relevant for simplicity. For example, in the $2d$ isotropic case, one may continue the Cartesian states or the polar states. If a basis state is complex-valued, then the real and imaginary parts are also valid linear states and should be considered for continuation if they generate new states.


We reiterate that this approach is essentially for understanding and benchmarking purposes, there is \textit{no} guarantee for convergence or existence, and it is not part of our systematic method. 
Our systematic linear limit continuation framework is closely related to the degenerate perturbation theory. It is extremely helpful to explain the motivation before presenting the theory and our solution to the degenerate state problem.

\textbf{Remark 2.} The nonlinearity significantly constraints the number of distinct stationary solitary waves bifurcating from a linear degenerate set, i.e., not all distinct linear states are continued to distinct nonlinear states in the near-linear regime. The number of distinct nonlinear wave patterns bifurcating from a finite linear degenerate set is well finite, and we call this phenomenon \textit{nonlinear locking} or \textit{nonlinear pinning}. Intuitively, this means that not all linear combinations yield distinct stationary states as a result of the nonlinear perturbation, only certain good linear combinations or wave patterns are self-consistently stationary.

This is a very important and even a key \textit{numerical fact} in the development of our theory, otherwise, the problem would be clearly intractable. At first sight, there is an infinite number of linear states to consider, as any linear combination of the linear degenerate states is also a perfectly valid linear stationary state. Fortunately, this is not the case. Here, we illustrate this fact by examining some low-lying states both theoretically and numerically. Consider the linear combination of the states \statem{10} and \statem{01} and focus now on real mixing, the node can only come from the polynomial part $\alpha x + \beta y=0$ as the background Gaussian profile has no node. This makes a rectilinear stripe profile passing through the center with only different orientations, there is no fundamentally new state. A less trivial example is the complex mixing, it is very well-known that the combination $(\state{10}+i\state{01})\sqrt{2}$ yields a vortex state. Both the combinations of $(\state{10}+2i\state{01})\sqrt{5}$ and $(2\state{10}+i\state{01})\sqrt{5}$ imprint vortex profiles as well, but they lead to the very same vortex state by the Newton's solver in the weakly nonlinear regime. After many similar trials, no deformed vortex state is found. This suggests that a very precise initialization of the initial guess may not be strictly necessary. In fact, this is likely why one can construct many solitary waves by physical insights, and then they are nevertheless continued successfully. Indeed, the continuation may still work even if the initial guess is only an approximation. 

The nonlinear locking can be understood in the framework of the degenerate perturbation theory. Consider a set of $K$ orthonormal linear degenerate states $\varphi_i^0, \ i=1, 2, ..., K$ with eigenenergy $\mu_0$. The degenerate perturbation analysis similarly works as:
\begin{align}
    \varphi^0 &=\sum_i c_i \varphi_i^0, \quad \sum_i|c_i|^2=1, \\
    \psi^0 &=\sqrt{\epsilon}\varphi^0, \\
    \mu &= \mu_0 + \epsilon \mu_r, \\
    \mu_r &= \iint |\varphi^0|^4 dx dy, \label{mur} \\
    \mu_rc_i &=\iint \varphi^2\varphi^*\varphi_i^*dxdy. \label{murci}
\end{align}
Here, $\epsilon$ is again a small perturbation parameter and the stationary superscript for the fields in the last equation is omitted for clarity.
The last equation is new, it comes from averaging over each individual linear basis state, and it gives $K$ constraints. In addition, if we multiply the equation by $c_i^*$ and then sum over $i$, we obtain the formula of $\mu_r$, therefore, Eq.~\eqref{mur} is not an independent constraint.
As a result, we end up with $K+1$ variables $\mu_r, \ c_i$ and $K+1$ constraints.
This explains why the number of solitary waves bifurcating from a finite linear degenerate set is finite. 
As the algebraic equations are nonlinear, it is evident that not any linear stationary state is expected to have a corresponding nonlinear stationary state. The nonlinearity is somewhat picky about the linear combination coefficients, leading to the nonlinear locking phenomenon. 

Importantly, we define the linear state $\varphi^0$ as the \textit{underlying linear state} (ULS) of the nonlinear state $\psi^0$ in the near-linear regime. The significance is that $\sqrt{\epsilon}\varphi^0$ is a good representation of the nonlinear wave $\psi^0$ at $\mu=\mu_0+\epsilon\mu_r$ to leading order in a reasonable interval in the near-linear regime. This is important for both a reliable theoretical analysis of a nonlinear wave, as well as for providing a good initial guess to a numerical solver. As we shall see later, the theoretically constructed ULS by physical insights can be oversimplified, while the true ULS can be quite complicated. In this setting, theoretical analysis can be combined with the numerically exact ULS in a semi-analytical way. This can be crucial for computing the properties of some solitary waves correctly in the small-density regime \cite{Wang:RDS}. The ULS was implicitly applied therein, here, we clearly point out its theoretical significance. Therefore, as we identify distinct nonlinear waves in this work, we map out their ULSs in detail. 

We aim to directly find numerically exact solutions in the near-linear regime at the PDE (partial differential equation) level. While it is possible to find distinct solitary waves by solving the algebraic equations, this approach is clearly rather tedious and elaborate. Our idea is reasonable as we need to continue the states into the TF regime using the PDE approach anyway. First, we find a solution $\psi^0$ at a chosen $\mu \gtrsim \mu_0$. If we normalize the state, we can extract its corresponding ULS $\varphi^0$. Furthermore, we can readily compute the numerically exact linear combination coefficients $c_i$ by expanding $\varphi^0$ on the basis states. Now, we are ready to discuss how to efficiently construct solitary waves from a given linear degenerate set.

The finite number of solitary waves bifurcating from a finite linear degenerate set motivates us to design a random searcher for finding distinct solitary waves in the near-linear regime. The initial state is prepared using a suitable random linear combination of the basis states. This turns out to be a very powerful and effective method with excellent convergence properties, and it is much more robust than we expected in the early stage of this work. We should emphasize that this is also a very important and key \textit{numerical fact} in the development of our theory, it is a property of the Newton's solver, independent of the degenerate perturbation theory. The excellent convergence property is most likely because the nonlinearity is sufficiently weak and only these few modes are essentially relevant in our setup. If a distinct solitary wave is found, we can calculate its ULS and the numerically exact linear combination coefficients, despite that the coefficients of the initial guess may be far from perfect. 

The random searcher also approximately gives us a sense of the completeness of the solitary waves bifurcating from a linear degenerate set. If a random solver is suitably designed to effectively explore the linear combination coefficients space, and we keep getting the same set of solitary waves after an increasing number of independent runs, it is like that no other solitary waves are expected. A physical picture is that these solitary waves are the fixed points in the linear combination coefficients space, and each state has its own basin of attraction. As the Newton's solver has a good convergence property, it is likely that each solitary wave has a reasonably good basin of attraction. Our random solver can find all the solitary waves that we expected from a linear degenerate set by physical insights, and importantly it provides (much) more stunning solutions that we could hardly imagine.




\begin{algorithm}[t]
\caption{A random searcher of real solitary waves}  
\label{RRS}                       
\begin{algorithmic}
\REQUIRE A set of $K$ real basis states, number of trials $N_T$.
\ENSURE Real solitary waves near the linear limit at $\mu \gtrsim \mu_0$.
\FOR{$run=1:N_T$}      
\STATE Generate a random integer $M=2, 3, ..., K$, following the uniform distribution.
\STATE Choose randomly $M$ states, and prepare an initial guess by a linear combination of the $M$ states, the coefficient of each state is chosen randomly as $2rand()-1$. Normalize the coefficients, and scale the initial state following the perturbation theory at $\mu$.
\STATE Solve using the Newton's solver.
\IF{Converges}       
\STATE Compute the underlying linear state following the perturbation theory, and compute the linear combination coefficients.
\STATE Save the solution, the initial and the computed coefficients.
\ENDIF
\ENDFOR
\STATE Identify the distinct solitary waves, and continue these states further into the TF regime.
\end{algorithmic}
\end{algorithm}

\begin{algorithm}[t]
\caption{A random searcher of complex solitary waves} 
\label{CRS}                       
\begin{algorithmic}
\REQUIRE A set of $K$ real basis states, number of trials $N_T$.
\ENSURE Complex solitary waves near the linear limit at $\mu \gtrsim \mu_0$.
\FOR{$run=1:N_T$}      
\STATE Generate a random integer $M=1, 2, 3, ..., K$, following the uniform distribution.
\STATE Choose randomly $M$ states, and prepare a real initial guess $u_r$ by a linear combination of the $M$ states, the coefficient of each state is chosen randomly as $2rand()-1$.
\STATE Repeat the previous two steps and prepare independently another real state $u_i$. The total initial guess is $u_0=u_r+iu_i$. Normalize the coefficients, and scale the initial state following the perturbation theory at $\mu$.
\STATE Solve using the Newton's solver.
\IF{Converges}                  
\STATE Compute the underlying linear state following the perturbation theory, and compute the linear combination coefficients.
\STATE Save the solution, the initial and the computed coefficients.
\ENDIF
\ENDFOR
\STATE Identify the distinct solitary waves, and continue these states further into the TF regime.
\end{algorithmic}
\end{algorithm}

We have designed two random searchers, a real random searcher (RRS) for purely real states, and a complex random searcher (CRS) for both real and complex states. The motivation of the RRS was historical, when we tried to construct complex states by a complex mixing of two real states. As such, we shall not discuss this further here. Now, we merely separate the two cases for clarity, and also for benchmarking purposes. Naturally, the CRS should find all the states found by the RRS. In addition, the CRS should find the well-known states bifurcating from the linear degenerate set, see, e.g., the first remark above. Once the algorithms are well established, one can then focus on the CRS in the future.

Our RRS is summarized in Algorithm~\ref{RRS}. First, we choose a number $M$ randomly, keeping $2\leq M \leq K$, where $K$ is the degeneracy. Then, $M$ basis states are selected randomly and their linear combination coefficients are taken as $2rand()-1$, where the random number $rand()$ follows the uniform $U[0,1]$ distribution. The initial guess is then suitably normalized at the working $\mu \gtrsim \mu_0$ following the perturbation theory, and it is subsequently applied to the Newton's solver for convergence. If the search converges, a solitary wave is found, and its ULS can be computed. This process can be repeated for $N_T$ times to exhaust the distinct wave patterns bifurcating from the linear degenerate set. We shall present more technical details soon, now we continue the discussion of the CRS.

Our CRS is similar in structure to the RRS and it works with a real basis set as well, as summarized in Algorithm~\ref{CRS}. If a basis set has complex states, one can readily construct a real basis set from the real and imaginary parts to apply the CRS. Basically, we apply the RRS twice for the real and imaginary parts of the initial guess, respectively. However, it is noted that $M=1$ is allowed in the CRS. The remaining parts are essentially the same. It is also valid to include $M=1$ in the RRS, then it is sufficient to implement the CRS in practice, as the RRS can be simply obtained by taking the real part of the initial guess. In addition, it is helpful to seed the random number generator and save the seed as well in the simulation. Finally, both algorithms are massively parallel due to the independent runs.

The choice of $\mu \gtrsim \mu_0$ is reasonably flexible, as the coefficients of the ULSs estimated are not very sensitive to the precise value of $\mu$ in the vicinity of the linear limit, in line with the perturbation theory. Clearly, it should not be too close to $\mu_0$ for numerical accuracy, note that the norm of the field exactly vanishes at the linear limit. On the other hand, it should not be unnecessarily too large such that the ULSs can be most faithfully represented by the pertinent basis states. Here, we work with $\mu=\mu_0+0.015$ unless otherwise specified. A helpful criterion is that the CRS should find all the known nonlinear states bifurcating from the linear degenerate set, as mentioned earlier. 

The convergence of a random search has two meanings. First, the solver successfully finds a stationary state. This appears to be very robust in our simulations. Second, this state should genuinely stem from the linear degenerate set of interest. In practice, the solver may occasionally converge to a state that bifurcated earlier from a smaller chemical potential. For example, when working with $\mu_0=2$, one may occasionally get the dark soliton stripe state, which bifurcates from $\mu_0=1$ but nonetheless it exists at $\mu=2$ as well. Fortunately, such events are not very common and they are also straightforward to identity as such states typically have ``anomlously'' large densities. By contrast, a solution that genuinely bifurcates from this linear set has a quite faint density as we are working with $\mu \gtrsim \mu_0$ in the near-linear regime. In this work, the criterion $|\psi^0|_{\rm{max}}<1$ is sufficient to filter the ``anomlous'' solutions. They can also be readily identified from the estimated coefficients. If a state truly stems from this linear degenerate set, then the numerical total weight $\sum_i |c_i|^2=1$ is satisfied to a good accuracy. Otherwise, the weight would be substantially smaller as the basis states are no longer complete for the state. 

Identifying distinct solitary waves can be largely done by observation, however, it is helpful to define a few statistics to be more reliable as some distinct solitary waves may look remarkably similar. To this end, we compute some statistics of $\psi^0$, e.g., $I_2=\iint |\psi^0|^2dxdy$, $|\psi^0|_{\rm{max}}$ and $|\psi(0)|$ for each converged state. These statistics are the norm of the field, the maximum magnitude, and the central magnitude, respectively. Importantly, they are invariant upto the symmetry transformations. Therefore, a joint plot of the triplet significantly simplifies the identification of distinct solitary waves.

Contour analysis plays a crucial role in understanding the nature of complex solitary waves. In this method, we plot the contours of both Re$(\psi^0)=0$ and Im$(\psi^0)=0$. The overlapping contours correspond to the dark soliton filaments, the intersection points signify the vortices. In addition, the relative charges of the vortices can also be readily extracted from these contours. The method is significant when several vortices cluster close together, generating a merged density depletion region and a corresponding complicated phase profile. To our knowledge, it can be extremely challenging to understand the nature of such solitary waves if one merely inspects the density and phase profiles. The method is by no means limited to the near-linear regime, it works equally well into the TF regime. As such, it also plays an important role in analyzing the chaotic processes of the chemical potential evolution, e.g., the vortex nucleation and pair creation, see Sec.~\ref{cc}.

Our solvers are by no means designed with any particular optimization, yet they appear to be reasonably effective and efficient. Therefore, it is likely that other reasonable designs also work well. For example, one may skip the selection of $M$ states and directly mix all the $K$ states with the random coefficients, either real or complex ones. In this setting, one may also directly work with a complex basis for the CRS. Then, the algorithm works in any linear basis of interest. Gaussian random numbers are also readily available. Future work should compare their efficiencies by analyzing the relative frequencies of visiting the different states.

Our numerical setup is now essentially complete. First, the linear degenerate sets are classified using lattice planes for the harmonic potential. This enables us to work with the linear degenerate sets in turn. For each degenerate set, we find distinct solitary waves bifurcating from it using the random searchers in the near-linear regime. Next, these states are further continued in the chemical potential into the TF regime. Once a state is in the TF regime, it may also be continued into other trap potentials. This program is clearly systematic, and also appears to be highly effective and efficient. In addition, it is massively parallel as the different linear degenerate sets and their solitary waves can be explored independently. There is one complication due to the presence of continuation chaos in $2d$, which we discuss in the next. However, this is largely due to the nature of solitary waves, rather than a limitation of our numerical setup.


\subsection{Continuation chaos}
\label{cc}
The continuation of many solitary waves in the chemical potential is highly robust, e.g., the ground state, the dark soliton stripe, the square vortex quadruple, and the polar states. However, the continuation of the others can be more complicated and even quite intriguing and complex. A solitary wave may evolve dramatically upon increasing the chemical potential, effectively morphing into a different structure. We refer to such transformations as \textit{chemical potential chaos}. Interestingly, the initial and final states may look strikingly different, even if the transition is a gradual and smooth crossover. Finally, a solitary wave may morph more than once from the near-linear regime to the TF regime.

There are many types of chemical potential chaos. We have identified and summarized here the following typical chaotic processes in $2d$ for dark soliton filaments and vortices. While we discuss below the chemical potential chaos, they are also relevant when we change the other system parameters such as the trap aspect ratio.
\begin{enumerate}
    \item Dark soliton deformation. A dark soliton filament may deform upon changing the chemical potential. A closed dark soliton loop extending out of the condensate may be induced into the condensate upon increasing the chemical potential.
    \item Dark soliton connection, disconnection, and reconnecttion.
    \begin{enumerate}
        \item Dark soliton connection. Two dark soliton filaments connect on their ends to form a closed dark soliton loop, which is typically noncircular. They can also connect by creating a single node, which is locally similar to the XDS2 structure.
        \item Dark soliton disconnection. This is the reverse process of the dark soliton connection, e.g., a closed dark soliton loop can break into two disconnected dark soliton filaments.
        \item Dark soliton reconnection. In this process, two dark soliton filaments come closer and exchange half of their filaments.
    \end{enumerate}
    \item Vortex reconfiguration. As the density grows, vortices may adjust their equilibrium configuration and thereby morph into a new structure.
    \item Vortex nucleation.
    \begin{enumerate}
        \item Single vortex nucleation. A single vortex can nucleate in a density depletion region. Particularly, an edge vortex frequently nucleates and subsequently be induced into the condensate to heal an edge density defect.
        \item Pair creation. A vortex-anti-vortex pair can directly nucleate in a density depletion region. Charge is conserved in this process.
        \item Elongation and pair creation. A vortex core may elongate and nucleate a total of three vortices of alternating charge, counting the original one. The total charge is conserved, but the central vortex charge is changed in this process.
    \end{enumerate}
    \item Vortex denucleation. Occasionally, a vortex may leave a density depletion region, but this process is typically followed by a pair creation to heal the density defect.
    \item Vortex merging. Several vortices in the condensate may be pushed together, forming effectively a single vortex of one or multiple charges. The vortices may have the same charge or mixed charges, and the total charge is conserved in this process.
    \item Pair annihilation. This is similar to the vortex merging, but a vortex-anti-vortex pair is pushed together, leaving a density depletion region with no vortex.
    \item Existence boundary. Sometimes, a solitary wave may cease to exist beyond a critical chemical potential $\mu_c$.
\end{enumerate}


Chemical potential chaos manifests either as a continuous gradual crossover or a discontinuous sudden transition in the solitary wave structure. In a crossover chaos, a state gradually morphs into a different structure with a series of intermediate states. By contrast, a transition chaos happens in a single step without the intermediate stationary states. It seems that a state simply ceases to exist beyond a critical chemical potential $\mu_c$, it undergoes a more dramatic structural change and directly morphs into a new state. Because of its similarity to a discontinuous phase transition in statistical mechanics, we also refer to it as a discontinuous state transition.

It is important to discuss the meaning to assign that the two pertinent states of a discontinuous transition chaos are parametrically connected. The motivation is that a series of quasi-intermediate states are provided by the Newton's solver. Two different interpretations are possible here. First, this ``evolution'' is somewhat real such that the initial and final states are weakly parametrically connected. Second, one may entirely reject this weaker connection for simplicity and conclude that the solitary wave no longer exists. We adopt the former perspective here if relevant for completeness to find more solitary waves. While the convergent process may be somewhat heuristic, it may indeed provide new and reasonably related structures. Nevertheless, we always state clearly whenever this happens, such that a reader who is not willing to establish such a connection can readily ignore the final state and its further continuation.

Chemical potential chaos slows down the numerical continuation, and the continuation of different states should be done in a case by case manner. A strongly chaotic state is typically much more expensive to continue than a state without chemical potential chaos. Dark soliton filaments are typically more flexible to continue than vortical patterns. A crossover chaos is also more computationally friendly than a transition chaos. Transition chaos presents a significant challenge to the continuation of solitary waves. A signature of transition chaos is that the continuation may fail no matter how small we choose the continuation step size $\delta \mu$ in practice. The solver typically relaxes to a much simper state, e.g., the ground state after a rather heuristic evolution. In this setting, we find that it is better to directly choose a moderate step size $\delta \mu$ such that the transition is realized ``in a single step''. In practice, this special step size $\delta \mu$ can be selected by trial and error. In our work, we have succeeded in capturing a successfully transition for each discontinuous transition chaos. It is an interesting question whether this is a generic feature. 

Chemical potential chaos is largely due to the intrinsic nature of some solitary waves, rather than a limitation of the linear limit continuation method nor numerical accuracy. Indeed, most chemical potential chaos, particularly transition chaos, occur in the intermediate regime or the TF regime. These chaos are relevant whenever the states are continued in the system parameter space, regardless where the continuation starts from. Chaos is also not a numerical illusion. For a genuine chemical potential chaos, it seems that it cannot be eliminated by merely increasing the numerical efforts, e.g., a finer spatial grid size $\delta x$ or a finer continuation step size $\delta \mu$. As such, it seems likely that chaos is intrinsic to the nature of the pertinent solitary waves.

\section{Results}
\label{results}

\subsection{Polar states and associated polar states}
\label{PS}

First, we emphasize that this section is motivated by the first remark. It is included here because it provides physical insights as the waves have beautiful patterns, and also for benchmarking purposes. This method is \textit{not} part of our systematic program, and indeed all the states herein are expected to appear automatically in our systematic search. The (associated) polar solitary waves are obtained by continuing the (real or imaginary parts of the) individual polar basis states. 

\begin{figure*}
\includegraphics[width=\textwidth]{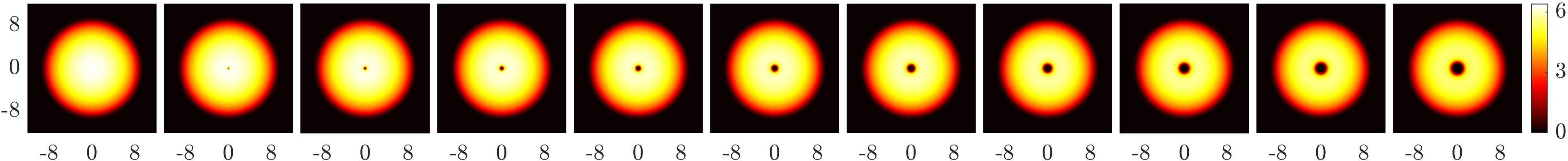}
\includegraphics[width=\textwidth]{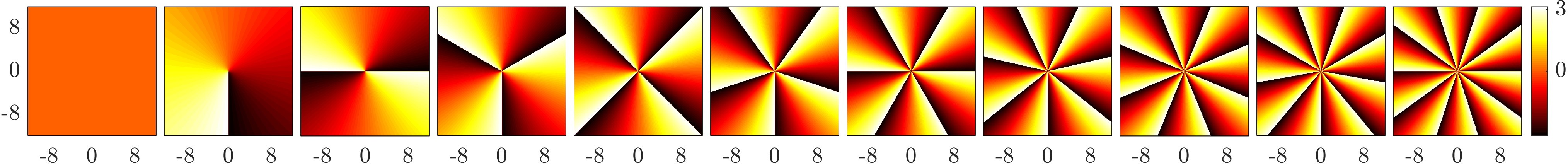}
\includegraphics[width=\textwidth]{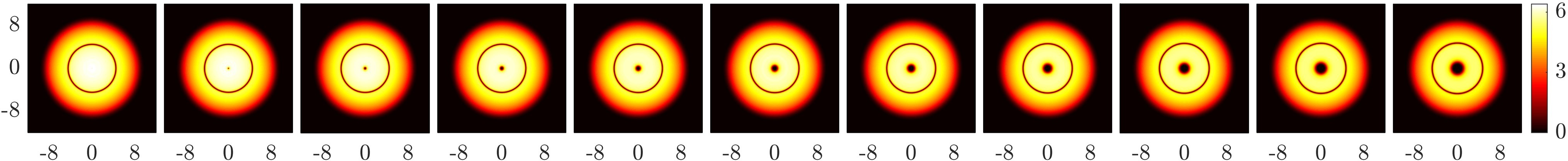}
\includegraphics[width=\textwidth]{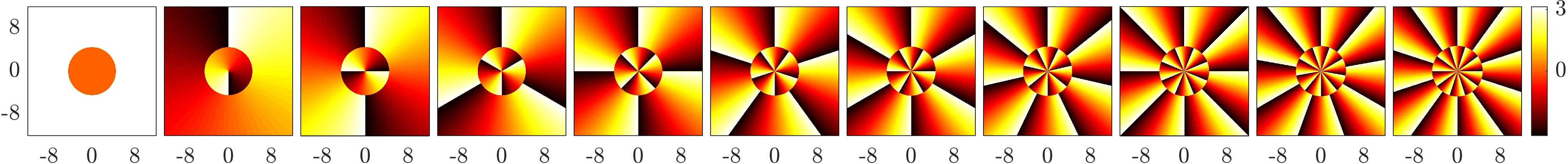}
\includegraphics[width=\textwidth]{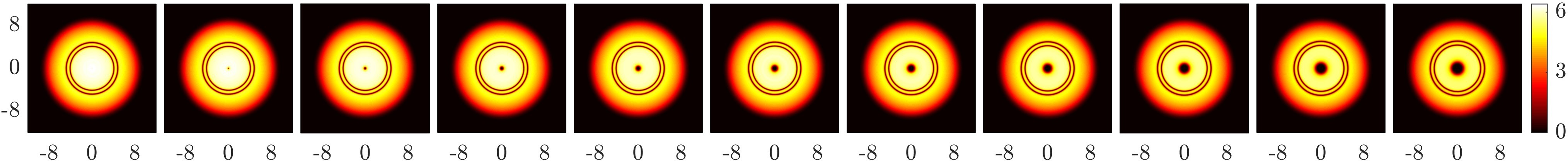}
\includegraphics[width=\textwidth]{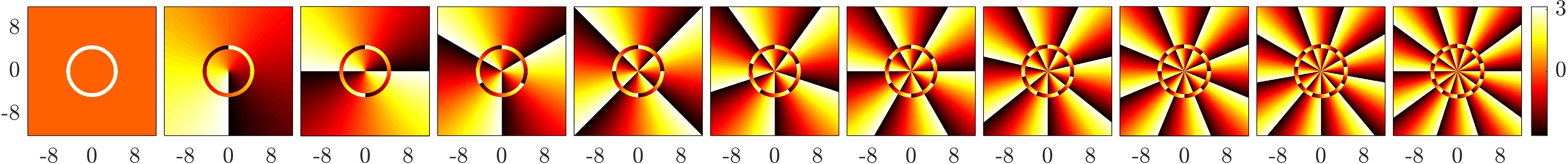}
\includegraphics[width=\textwidth]{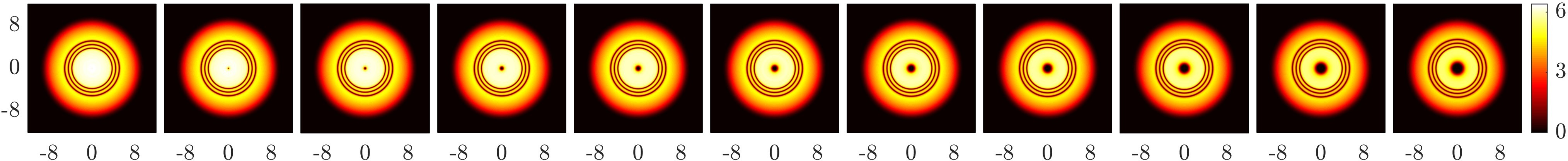}
\includegraphics[width=\textwidth]{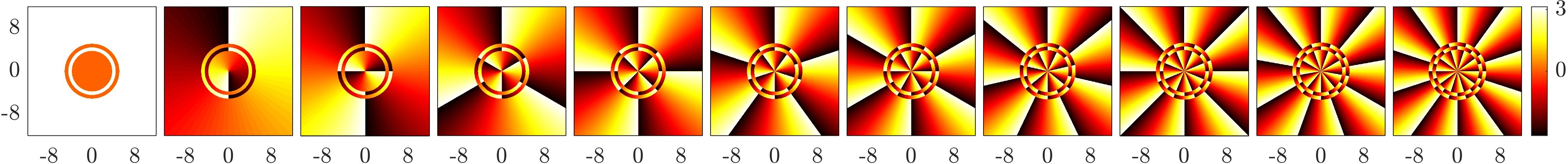}
\caption{
Typical solitary waves continued from the polar linear states, the odd and even panels depict the magnitude and phase profiles at $\mu=40$, respectively. The first set illustrates the low-lying states with no radial or ring node, from the left to right, they are the ground state, vortex states with topological charges $1, 2, 3, ..., 10$, respectively. The second set is similar, but the corresponding states have an additional RDS. Similarly, the corresponding states of the third and fourth sets have two and three RDSs, respectively. More complex states are possible, but the pattern is clear and hence they are omitted here for clarity.
}
\label{polarstates}
\end{figure*}

The continuation of polar states in the $1d$ framework, see Eq.~\eqref{GPE1d}, appears to be fully robust. They can be readily continued upto $\mu=80$ which is deep in the TF regime. A large set of low-lying states at $\mu=40$ is depicted in Fig.~\ref{polarstates}. These $2d$ states are obtained from their $1d$ counterparts by a cubic spline interpolation. These spectacular states exhibit a very clear pattern inherited from the linear states, each linear state \statelp{nm} is continued into a solitary wave \statep{nm} with $n$ concentric RDSs and a central vortex of charge $m$. In our computation, we have exhausted all the solitary waves of $0\leq n, m \leq 10$. In addition, we have continued a few randomly selected more excited states \statep{12,23}, \statep{17,5}, and \statep{24,12} and the continuation remains very robust. 

We expect that each linear polar state is genuinely the ULS of its own corresponding nonlinear state. This is because a nonlinear polar state has a well-defined topological charge as $\psi^0=u_n^0(r)\exp(im\theta)$. For a polar linear degenerate set, each basis state has a unique topological charge. Therefore, the ULS of $\psi^0$ can only come from the linear basis state of the same charge $m$ and it is orthogonal to all the other linear basis states of different charges. 
\begin{align}
    \varphi_{\mathrm{RDSnVX^m}}^0 &= \state{nm}_P.
\end{align}
This theoretical expectation is numerically confirmed for all the polar states studied above. Therefore, there is a simple one-to-one correspondence between a linear polar state and its nonlinear polar wave, a fact that is crucial for applying the continuation method to polar states in earlier works. It is important not to take this correspondence for granted, we shall see later that it does \textit{not} hold for the Cartesian basis states.

These polar states have been extensively studied in the past, including systematic efforts focusing on a few low-lying wave patterns \cite{Carr:VX,Wang:RDS2}.
The ground state and the single vortex state of charge $1$ appear to be fully robust, and all the other states suffer from various unstable modes at least in certain chemical potential intervals. The RDS is unstable right from the linear limit through the transverse instability, leading to the nucleation of vortex ``multipoles" as vortex squares, vortex hexagons, octagons, decagons, and so on \cite{Wang:DSR,Wang:AI}. A central vortex with a higher charge is energetically unfavourable, and it is typically unstable towards a spontaneous disintegration. Oscillatory instability arises in a quasi-periodic manner as the chemical potential increases, and the number of unstable modes tends to grow with increasing charge \cite{Law:VX2,Wang:RDS2,Panos:book}. The interactions between a central vortex and a RDS, and between multiple RDSs are also considered \cite{Wang:RDS2}. Naturally, the BdG spectrum typically becomes increasingly complex as the number of rings and the vortex charge increase, featuring more unstable modes. Nevertheless, it seems likely they can be fully stabilized with suitable external potential barriers placed at the radial density nodes, at least for the low-lying states \cite{Wang:RDS,Wang:RDS2}.

\begin{figure*}
\includegraphics[width=\textwidth]{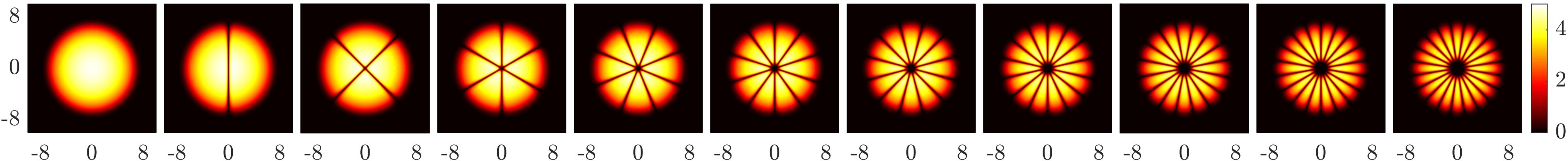}
\includegraphics[width=\textwidth]{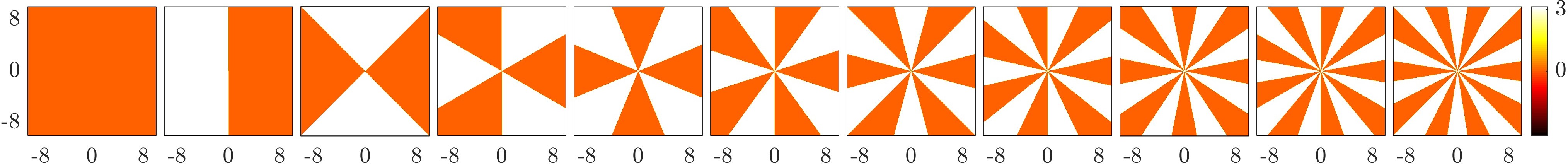}
\includegraphics[width=\textwidth]{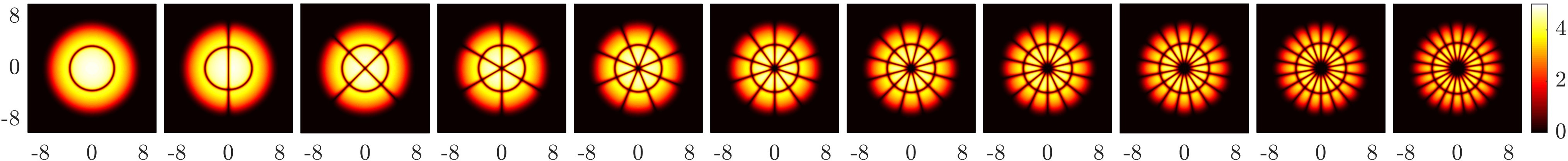}
\includegraphics[width=\textwidth]{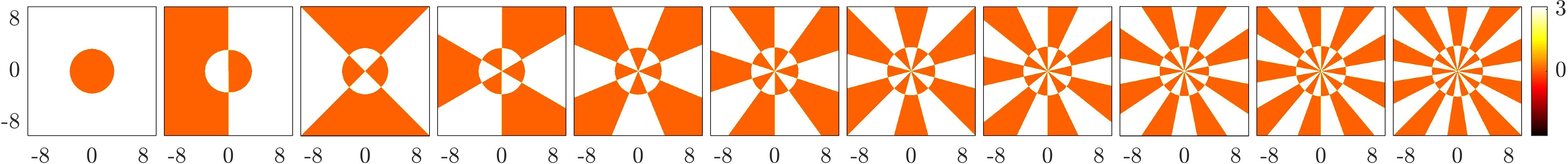}
\includegraphics[width=\textwidth]{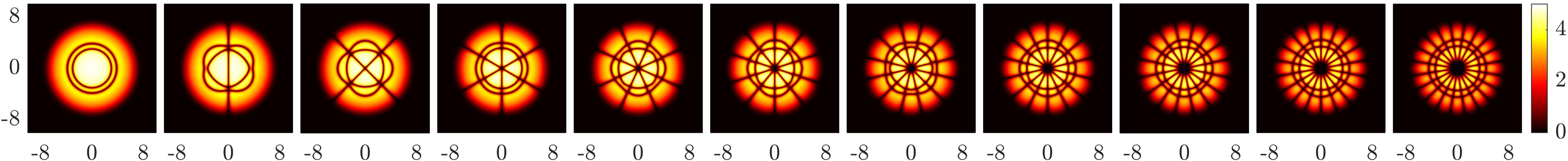}
\includegraphics[width=\textwidth]{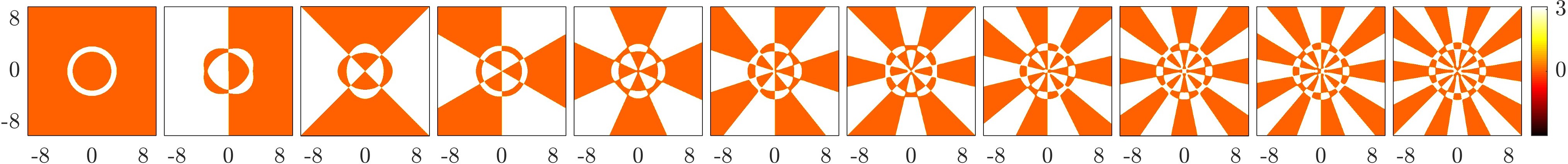}
\caption{
Associated polar states continued from the real part of the linear polar states \statelp{nm}, the depicted states are at $\mu=24$. The linear polar states correspond to the ones shown in Fig.~\ref{polarstates} but only the states upto two RDSs are illustrated.
}
\label{polarstates2}
\end{figure*}

The polar basis also offers new series of solutions from their real or imaginary parts, we refer to them as the associated polar states and some low-lying ones are depicted in Fig.~\ref{polarstates2}. Such states are anisotropic, and therefore should be solved in the genuine $2d$ framework. It is sufficient to focus on the real part, as the state from the imaginary part is identical upto a rotation. It is understood that a factor of $\sqrt{2}$ is applied to suitably normalize the linear states. As such, the central vortex of charge $m$ yields $m$ equally-spaced angular nodes passing through the origin by the $\cos(m\theta)$ function. We refer to this structure as the cross dark soliton XDS$_m$, also known as multi-poles in the terminology of \cite{RCG:rotating}.
The RDS profile remains unchanged upon taking the real or imaginary parts. Therefore, the associated polar state stemming from the linear polar state \statelp{nm} is the RDSnXDSm with $n$ concentric RDSs and $m$ XDSs in the ideal case. However, the ULSs of the associated polar states are typically quite complicated.

It seems likely that the linear states obtained above are genuine ULSs for the XDSs without any RDS, but typically not for the solitary waves with RDSs, they are only approximately so. The XDSs appear to be equally spaced throughout the continuation, and numerical results strongly suggest that:
\begin{align}
    \varphi_{\mathrm{XDSm}}^0 &= \frac{1}{\sqrt{2}}(\state{0,m}_P+\state{0,-m}_P).
\end{align}
This is numerically confirmed for all the XDSs of $0<m\leq 10$. The ULSs of the associated polar states with rings are more complicated, because the RDSs can become (strongly) distorted in the near-linear regime. For example, the $\phi$ soliton has a single dark soliton stripe and a single RDS, the RDS takes an elliptical shape in the weakly-interacting regime. The nonlinearity can have a significant impact on the equilibrium configuration, e.g., the ring appears to get increasingly circular as the chemical potential increases. The $\phi$ soliton therefore is like a simple composition of a RDS and a stripe dark soliton in the TF regime. However, the two rings in the RDS2XDS and RDS2XDS2 states are significantly distorted in both the near-linear regime and the TF regime. Because of this complexity, we shall discuss these states one by one as we encounter them in our systematic search. There is also an interesting analogy between the dark soliton filaments herein and the vortical filaments of the Chladni solitons found in \cite{Chladni}. Intuitively, one can extend the dark soliton filaments along the $z$ axis to dark soliton surfaces in a $3d$ cylindrical condensate. Then, the low-lying Chladni solitons are expected to approximately arise from the complex mixing of the extended $3d$ dark soliton surfaces and the single dark soliton plane \statem{001} in a suitable cylindrical trap.


\subsection{States from the isotropic trap}
\label{SST}

For the isotropic trap, one can work with either the Cartesian basis or the polar basis. Here, we work with the real Cartesian basis, as it is more generic for the trap aspect ratio. We now explore the low-lying linear degenerate sets in turn. The corresponding lattice planes are depicted in Fig.~\ref{XY}.

The lowest set is the ground state \statem{00} with eigenenergy $\mu_0=1$, and its continuation as a function of the chemical potential is already shown in Fig.~\ref{GS}. This state is nondegenerate, and it is also the ground state in the polar basis. Naturally, the ULS in the near-linear regime is:
\begin{align}
    \varphi_{\mathrm{GS}}^0 &= \state{00}.
\end{align}
The ground state takes a Gaussian profile in the near-linear regime. The continuation appears to be fully robust and we have continued the ground state upto $\mu=80$, deep in the TF regime. In the TF regime, the ground state takes a Thomas-Fermi profile:
\begin{align}
    \psi_{\mathrm{TF}}^0 &\approx \sqrt{\max(\mu-V,0)}.
\end{align}
The size of the condensate is characterized by the TF radius $R_{\mathrm{TF}}=\sqrt{2\mu}/\omega$, and the maximum density is approximately $|\psi_{\mathrm{TF}}(0)|^2=\mu$ at the trap center where $V=0$.
The ground state has a uniform phase, and it provides the background for hosting the embedded solitary waves. The BdG spectrum of the ground state in the harmonic trap is well known, see, e.g., \cite{Wang:RDS2,PK:GS}. It is fully robust and stable at all chemical potentials.


The second linear degenerate set includes two states \statem{10} and \statem{01} with eigenenergy $\mu_0=2$. Our RRS and CRS each with $100$ runs find two distinct states, a dark soliton stripe and a single vortex state. The ULSs are, respectively,
\begin{align}
    \varphi_{\mathrm{DS10}}^0 &= \state{10}, \\
    \varphi_{\mathrm{VX}}^0 &= \frac{1}{\sqrt{2}}(\state{10}+i\state{01}).
\end{align}
Their continuations are also very robust, as illustrated in Fig.~\ref{GS}. We have also successfully continued both states upto $\mu=80$, deep in the TF regime. They are also found in the polar and the associated polar states. 

The dark soliton state has a rectilinear density node passing through the origin, featuring a phase jump of $\pi$ across the node. The single vortex sits at the trap center, with a long-range phase winding around a short-range density core. The complex conjugate of the vortex state corresponds to an antivortex state of charge $-1$. As the superfluid velocity is proportional to the phase gradient of the wavefunction, the condensate flows counterclockwise and clockwise for the two vortices, respectively. 
The dark soliton stripe and the single vortex have both been extensively studied. The dark soliton BdG spectrum can be found in \cite{PK:DSVX}.
It is stable only when it is sufficiently close to the linear limit, and it quickly becomes increasingly unstable as $\mu$ increases. It also suffers from the well-known transverse instability, leading to the nucleation of vortex-anti-vortex pairs \cite{Ma:DS}.
By contrast, the vortex is highly robust and it is fully stable at all chemical potentials, see, e.g., \cite{PK:DSVX,Wang:RDS2} for the BdG spectrum. The vortex precesses when it is shifted off the trap center, and the precessional frequency is $\omega_{pr}=\omega^2/(2\mu)\log(A\mu/\omega)$, where $A\approx 8.88$ \cite{PK:DSVX,Wang:VX}.

\begin{figure*}
\includegraphics[width=\textwidth]{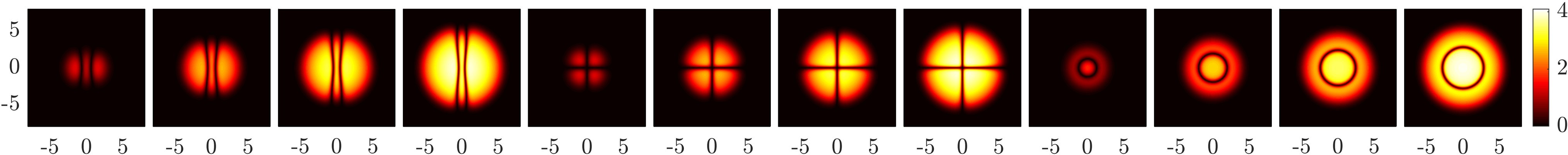}
\includegraphics[width=\textwidth]{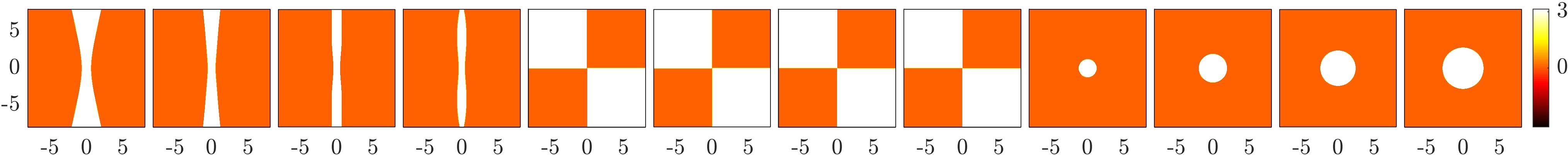}
\includegraphics[width=\textwidth]{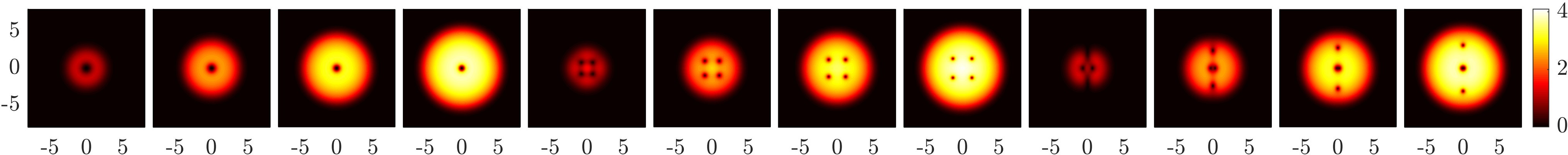}
\includegraphics[width=\textwidth]{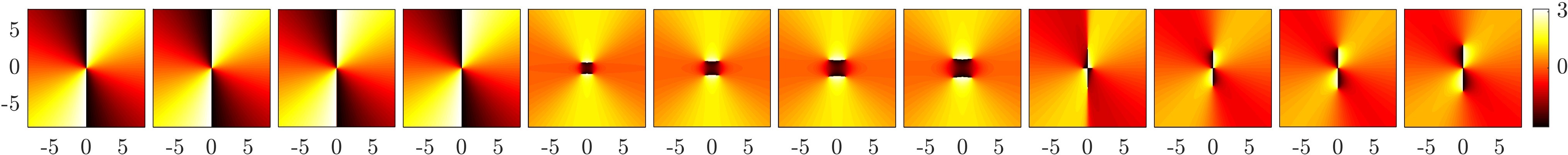}
\caption{
Solitary waves continued from the linear degenerate states of $\mu_0=3$, these states are the DS20, DS11, RDS, VX$^2$, SVQ, and RVQ, respectively. Here, typical states at $\mu=4, 8, 12, 16$ are depicted for each wave configuration. Particularly, the RVQ undergoes a gradual crossover chaos, becoming an aligned vortex triple state in the TF regime.
}
\label{DS20}
\end{figure*}

The third linear degenerate set has a total of three basis states \statem{20}, \statem{11}, and \statem{02} with eigenenergy $\mu_0=3$. To gain more physical insights, it is helpful to know their wavefunctions.  Upto a common normalization factor and the ground state background, the pertinent polynomial parts are $(2x^2-1)/\sqrt{2}, 2xy$, and $(2y^2-1)/\sqrt{2}$, respectively. 
We have identified a total of $6$ distinct solitary waves from this degenerate set, as depicted in Fig.~\ref{DS20}. The ULSs are summarized as:
\begin{align}
    \varphi_{\mathrm{DS20}}^0 &\approx 0.9959\state{20}-0.0906\state{02}, \\
    \varphi_{\mathrm{DS11}}^0 &= \state{11}, \\
    \varphi_{\mathrm{RDS}}^0 &= \frac{1}{\sqrt{2}}(\state{20}+\state{02}), \\
    \varphi_{\mathrm{VX^2}}^0 &= \state{02}_P, \\
    \varphi_{\mathrm{SVQ}}^0 &\approx \frac{1}{\sqrt{2}}\state{20}+(0.0905+0.7013i)\state{02}, \label{SVQ}\\
    \varphi_{\mathrm{RVQ}}^0 &\approx 0.6739\state{20}+0.7373i\state{11})+0.0480\state{02}.
\end{align}
These coefficients are estimated using the numerically exact states at $\mu=3.015$, but they depend only very weakly on $\mu$ in the near-linear regime. The single vortex state of charge $2$ is expressed in the polar basis for simplicity. We have conducted $1000$ independent runs for both the RRS and the CRS for this degenerate set.

Interestingly, some of the linear combination coefficients are quite intriguing. For example, the DS20 state is mainly from \statem{20} but it also has a small contribution from \statem{02}, cf. the polar states. Consequently, the two dark soliton stripes are not strictly parallel, but rather they take a hyperbolic shape immediately in the near-linear regime, as shown in Fig.~\ref{DS20}. The hyperbolic equation is approximately $x^2/0.4545-y^2/4.9961=1$. Similar curving phenomenon is also found in the three-dimensional RDS, which is a hyperboloid of one sheet in the near-linear regime \cite{Wang:RDS}. 
This clearly illustrates that a theoretically constructed linear state may be only an approximation, and the effect of such subtle difference sometimes can be significant, showing the importance of the numerically exact ULSs \cite{Wang:RDS}. The DS20 state is stable when it is sufficiently close to the linear limit \cite{Middelkamp:VX}. In the TF regime, the dark soliton filaments take a more complicated shape and tend to form a closed loop. The continuation of the DS11, RDS, and VX$^2$ states appears to be fully robust. We obtained these states earlier in Sec.~\ref{PS}, and now we find them automatically in our random search. Both the DS11 and RDS states become unstable when departing from the linear limit towards forming vortical structures \cite{Middelkamp:VX,Wang:RDS2}. But full stabilization is possible for the RDS with an external potential barrier \cite{Wang:RDS,Wang:RDS2}, and it is likely that this also works for the DS11 state. The VX$^2$ spectrum is shown in, e.g., \cite{Wang:RDS2}, featuring quasi-periodic oscillatory unstable modes.

To gain more physical insights, we also present some theoretical analyses of manipulating the linear states. It is obvious that $(\state{20} \pm \state{02})/\sqrt{2}$ are interesting combinations, the addition $(\state{20}+\state{02})/\sqrt{2} \propto (r^2-1)$ yields the RDS while the subtraction $(\state{20}-\state{02})/\sqrt{2} \propto (x^2-y^2)$ gives a XDS2 oriented along the diagonals $y=\pm x$. This is a rotated $\state{11} \propto xy$ by $\pi/4$. 
The complex mixing of these two XDSs yields the vortex state of charge $2$. Alternatively, the two states are the real and imaginary parts of the polar VX$^2$ state. The complex mixing of \statem{20} and \statem{02} suggests a square vortex quadruple (SVQ) state, and the CRS also finds a rhombus vortex quadruple (RVQ) state \cite{Panos:DC1}. The theoretical analysis here is not required for applying our method, and there is no guarantee that the construction by physical insights is exact or even exists. For example, one may expect elliptical rings by mixing these second order polynomials, however, only the perfectly circular RDS is found in the weakly-interacting regime. This is just another example of nonlinear locking.

The ULS of the SVQ is somewhat complicated, it is \textit{not} a simple complex mixing as $(\state{20}+i\state{02})/\sqrt{2}$ as widely believed, despite that it is approximately so. For the state of Eq.~\eqref{SVQ}, the real part is an ellipse dark soliton and the imaginary part is the \statem{02} state.
It is remarkable that the vortices still form a perfect square by numerically solving $\varphi_{\mathrm{SVQ}}^0=0$. However, it should be noted that the real and imaginary parts of a complex state have \textit{no} absolute meaning, as they change with a global phase shift. Fortunately, the location and the charge of the vortices do not depend on this phase shift. Indeed, one can make a phase shift to the SVQ state such that the two coefficients are complex conjugate of each other. As mentioned earlier, the real and imaginary parts are then the RDS and the diagonal XDS2, respectively, despite the magnitudes of the coefficients are different. In this form, it is evident that the vortices of the ULS are located at $(\pm 1/\sqrt{2}, \pm 1/\sqrt{2})$, in line with the numerical analysis. The SVQ is found to be quite robust except for a narrow interval of oscillatory instability \cite{PK:DSVX,Middelkamp:VX}.

The four vortices have another equilibrium configuration, the RVQ, and its ULS is also somewhat complicated. It seems reasonable to call the SVQ and the RVQ solitary wave \textit{isomers}. Note that the vortex charge alternates along both quadrilaterals. Similarly, we can numerically calculate that the vortices of the ULS are located at $(\pm 0.7319,0)$ and $(0,\pm 2.7422)$. The RVQ is stable in a narrow chemical potential interval near the linear limit \cite{Panos:DC1}. Another isomer is known as the aligned vortex quadruple, the vortex charge also alternates, it has a linear limit in a suitable anisotropic trap but it also exists in the isotropic potential \cite{PK:DSVX}.


The RVQ exhibits chemical potential chaos as shown in Fig.~\ref{DS20}, despite that it is a relatively low-lying state. As the chemical potential increases, the two faint edge vortices are gradually induced into the condensate at about $\mu=8$. More importantly, the two inner vortices near the origin are pushed increasingly closer and they essentially merge together into a single vortex of charge $2$ at the center in the TF regime. The final state is therefore more like an aligned vortex triple state with relative charges $-1, 2, -1$. This state appears to be new and was not reported before, to our knowledge. 
Interestingly, the states in the near-linear regime and the TF regime are strikingly different at first sight. Nevertheless, this appears to be a gradual crossover rather than a sudden transition. This example also highlights the importance of a systematic continuation, otherwise, the link can be readily missed. 

\begin{figure*}
\includegraphics[width=\textwidth]{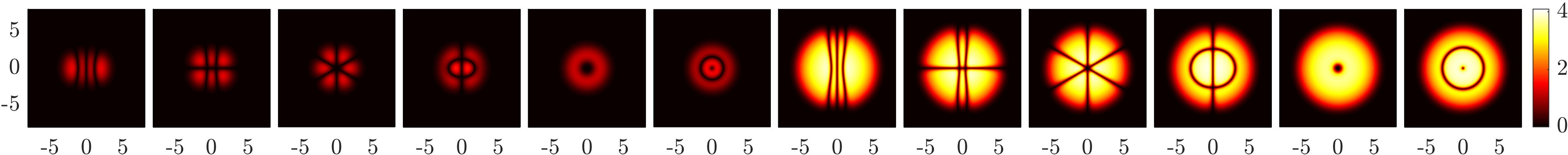}
\includegraphics[width=\textwidth]{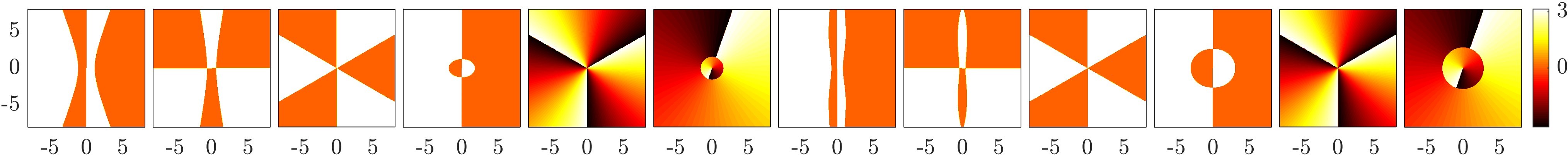}
\includegraphics[width=\textwidth]{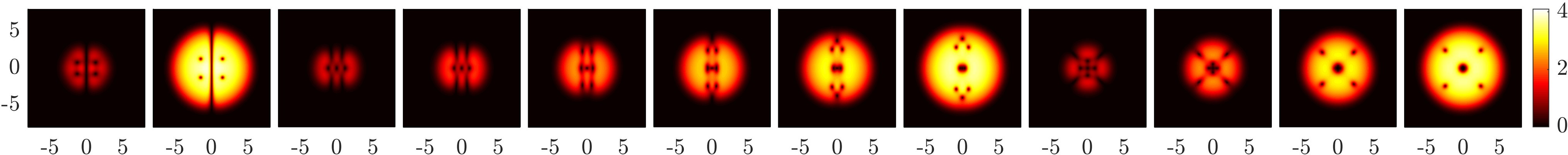}
\includegraphics[width=\textwidth]{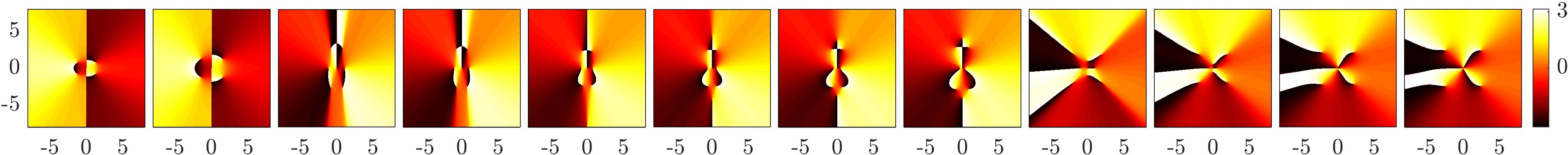}
\includegraphics[width=\textwidth]{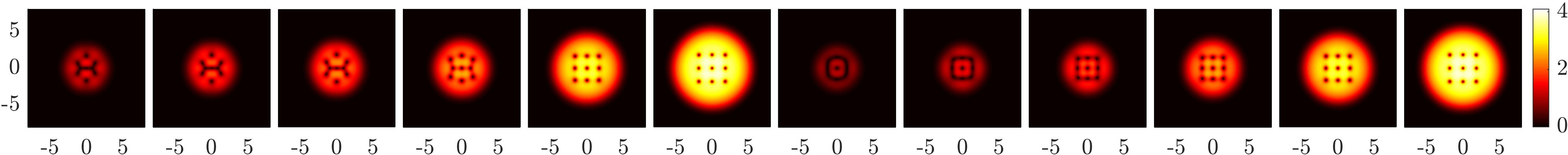}
\includegraphics[width=\textwidth]{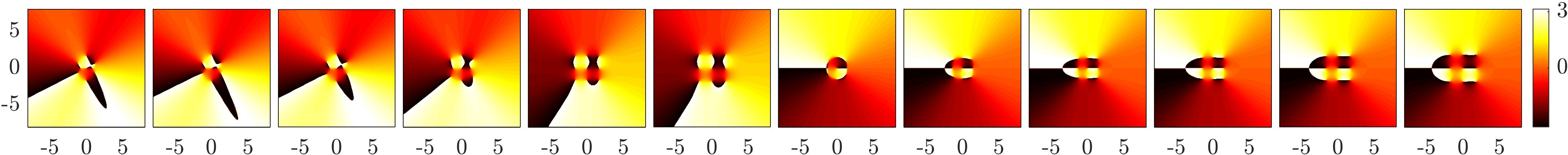}
\includegraphics[width=\textwidth]{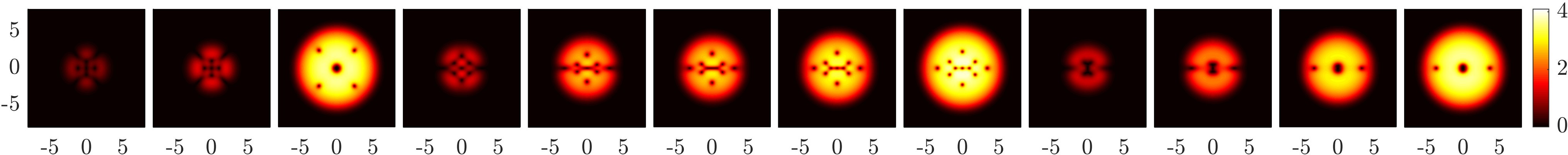}
\includegraphics[width=\textwidth]{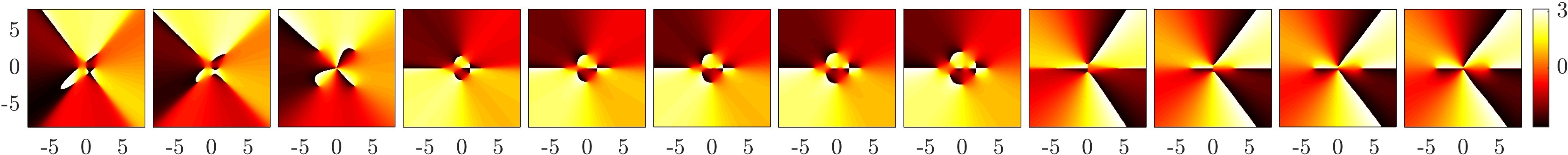}
\caption{
Solitary waves continued from the $\mu_0=4$ linear degenerate states, the top set depicts typical states at $\mu=5$ and $\mu=16$, respectively. The second set is for the DSVX4 state at $\mu=5, 16$, the VX7 state at $\mu=5, 6, 8, 10, 12, 16$, and the VX9a state at $\mu=5, 8, 12, 16$. The third set is for the VX9b state at $\mu=5, 6, 7, 8, 12, 16$, and the VX9c state at $\mu=4.5, 5, 6, 8, 12, 16$. The fourth set is for the VX9d state at $\mu=4.3, 5, 16$, the VX9e state at $\mu=5, 9, 10, 12, 16$, and the VX9f state at $\mu=5, 8, 12, 16$.
}
\label{DS30}
\end{figure*}

The fourth degenerate set has four basis states at $\mu_0=4$, and there is an explosion in the number of possible solitary waves as depicted in Fig.~\ref{DS30}. Indeed, some solitary wave structures are even quite stunning to imagine. 
The polynomial parts of the linear basis states upto a common factor are $(2x^2-3)x/\sqrt{3}$, $(2x^2-1)y$, $(2y^2-1)x$, $(2y^2-3)y/\sqrt{3}$ for the states $\state{30}, \state{21}, \state{12}, \state{03}$, respectively. First, we summarize the ULSs:
\begin{align}
    \varphi_{\mathrm{DS30}}^0 &\approx 0.9916\state{30}-0.1296\state{12}, \\
    \varphi_{\mathrm{DS21}}^0 &\approx 0.9938\state{21}-0.1112\state{03}, \\
    \varphi_{\mathrm{XDS3}}^0 &=\frac{1}{2}(\state{30}-\sqrt{3}\state{12}), \\
    \varphi_{\mathrm{Phi}}^0 &\approx 0.6417\state{30}+0.7670\state{12}, \\
    \varphi_{\mathrm{VX^3}}^0 &= \state{03}_P, \\
    \varphi_{\mathrm{RDSVX}}^0 &= \state{11}_P, \\
    \varphi_{\mathrm{DSVX4}}^0 &\approx 0.6684\state{30}+(0.0873-0.7387i)\state{12}, \\
    \varphi_{\mathrm{VX7}}^0 &\approx 0.6611\state{30}-0.7461i\state{21} \nonumber \\
    &+0.0469\state{12}+0.0632i\state{03}, \\
    \varphi_{\mathrm{VX9a}}^0 &\approx 0.6734(\state{30}+i\state{03}) \nonumber \\
    &-0.2157(i\state{21}+\state{12}), \\
    \varphi_{\mathrm{VX9b}}^0 &\approx 0.5367\state{30}-0.5456i\state{21} \nonumber \\
    &-0.3779\state{12}-0.5210i\state{03}, \\
    \varphi_{\mathrm{VX9c}}^0 &\approx 0.6440\state{30}+0.4675i\state{21} \nonumber \\
    &+0.0421\state{12}+0.6041i\state{03}, \\
    \varphi_{\mathrm{VX9d}}^0 &\approx 0.6630\state{30}-0.4731i\state{21} \nonumber \\
    &-0.1126\state{12}+0.5692i\state{03}, \\
    \varphi_{\mathrm{VX9e}}^0 &\approx 0.0836\state{30}-0.5848i\state{21} \nonumber \\
    &+0.6572\state{12}-0.4681i\state{03}, \\
    \varphi_{\mathrm{VX9f}}^0 &\approx 0.0101\state{30}+0.5860i\state{21} \nonumber \\
    &-0.7023\state{12}-0.4042i\state{03}.
\end{align}
These coefficients are estimated using the numerically exact states at $\mu=4.015$, and we have conducted $1000$ independent runs for both the RRS and the CRS. Here, numerical contour analysis depicting the contours of Re$(\psi^0)=0$ and Im$(\psi^0)=0$ becomes extremely helpful in analyzing the nature of some complicated states and their chaotic chemical potential evolution.


The DS30 and DS21 states are qualitatively similar to the previous DS20 state, with an additional vertical or horizontal dark soliton stripe through the center, respectively. Note also the relative minus signs in the wavefunctions of their ULSs. In the TF regime, the off-axis dark soliton filaments again take a more complicated shape and the filaments tend to form closed loops. The XDS3 was found previously from the real part of the VX$^3$ state, and it is remarkably an ULS. This is no longer the case for the $\phi$ soliton as mentioned earlier. The ULS is not merely the real part of the RDSVX state. The ring is an ellipse, and the dark soliton stripe sits along the minor axis. The equation of the ellipse is approximately $x^2/2.5351+y^2/1.2245=1$. However, it seems that the ring becomes increasingly circular as the chemical potential increases. 
We have also found the VX$^3$ and the RDSVX polar states as expected. The continuation of all the above states appears to be fully or reasonably robust. The DS30 is again stable when it is sufficiently close to the linear limit \cite{Middelkamp:VX}. The RDSVX is unstable right from the linear limit, but it can be suitably stabilized with external potential barriers \cite{Wang:RDS2}. The VX$^3$ can be fully stable, but it has two series of quasi-periodic oscillatory unstable modes \cite{Wang:RDS2}.

The DSVX4 state is a mixed state of a dark soliton stripe and four vortices of alternating charge. It is largely from the complex mixing of the \statem{30} and \statem{12} states, but the ULS is quite complicated. The vortices of the ULS sit at $(\pm 1.2247, \pm 0.7071)$, they take a rectangular shape rather than a perfect square in the near-linear regime. However, it seems that they evolve towards a SVQ in the TF regime. 

The VX7 state has a pretty dramatic evolution, because of its strong density depletion regions. It is largely from a complex mixing of \statem{30} and \statem{21} in the low-density regime. This explains the aligned vortex triple, all of charge $-1$, along the $x$ axis, and also the strong density depletion on the sides. The other basis states are mixed too, leading to four edge vortices of charge $1$ at the density depletion wedges. The four edge vortices are gradually induced into the condensate at about $\mu=8$. 
Interestingly, the density therein remains somewhat depleted and another two edge vortices of charge $-1$ are subsequently nucleated at about $\mu=8.1$, and they are also induced into the condensate at about $\mu=12$ and the condensate background finally takes an isotropic shape. These two vortices are not present in the ULS, suggesting that the number of building blocks of excitations can vary as a function of the chemical potential. Finally, the central three vortices are pushed together to form a single vortex of charge $-3$ at the center in the TF regime. This evolution illustrates again that the states in the near-linear regime and the TF regime can look strikingly different.

We have found a remarkable array of VX9 states, and they all show chemical potential chaos to various degrees. In the VX9a state, there is a central vortex of charge $1$ surrounded by four vortices of charge $-1$. They take a plus shape. In the diagonal directions, there are another four vortices of charge $1$ at the edge of the condensate. This state is qualitatively like a complex mixing of the DS30 and the DS03 states, see the real and the imaginary parts of its ULS. As the outer dark soliton stripes take a hyperbolic shape, the nine vortices take the configuration above rather than a square lattice in the near-linear regime. As the chemical potential increases, the four vortices are induced into the condensate at about $\mu=8$. In addition, the central $5$ vortices are pushed together, forming effectively a single vortex of charge $-3$ at the center in the TF regime. Interestingly, the vortex merging here involves a cluster of vortices of mixed charges. It was reported that both the VX7 and VX9a states appear to be oscillatorily unstable in the vicinity of the linear limit \cite{Panos:DC1}.

The VX9b state has seven vortices in the central part taking a dog bone shape with an additional vortex on each side in the near-linear regime. The seven vortex cores strongly overlap, forming a connected density depletion region inside the condensate. The vortical structure becomes evident at about $\mu=7$ where the density is moderately large and yet the state remains qualitatively similar to its ULS. It is interesting that the VX9b state gradually evolves into a $3\times3$ miniature square lattice, and the vortex charge alternates between the nearest neighbours. The states in the near-linear regime and the TF regime again look very different, and in fact \textit{no} such a square lattice is found in the low-density regime. Therefore, the ULS of the square lattice is not a simple complex mixing of the \statem{30} and \statem{03} states. This evolution is a gradual crossover rather than a sudden transition, the vortices slowly adjust their equilibrium positions towards a square lattice in the process. The crossover is most rapid between about $\mu=7$ and $\mu=8$. Note that the state at $\mu=8$ is like an interesting intermediate transition state.

The VX9c state has a central vortex surrounded by eight vortices featuring approximately an elongated ring structure in the small-density regime. The vortex cores again strongly overlap, leading to a connected density depletion racetrack. A detailed analysis of the ULS shows that there are two vortices along the $x$ axis and two vortices along the $y$ axis, each vortex along the $x$ axis is surrounded by another two vortices nearby. Therefore, the angular distribution of the vortices along the ring is not uniform. As the chemical potential increases, the vortices quickly become more prominent and remarkably also reorganize into the $3\times3$ square lattice; see the state at $\mu=6$. Despite the rather different evolutions, the square lattice in the TF regime appears to be identical, to our knowledge. Particularly, compare the states at $\mu=12$ and $16$. The evolution of the VX9b and the VX9c states suggests that different ULSs may evolve into the same nonlinear wave in the TF regime, i.e., the linear limit of a solitary wave is remarkably not necessarily unique. This miniature vortex lattice is known to be spectrally unstable, and it is also observed as a transient state in the decay of the RDSVX structure \cite{Middelkamp:VX}.

The VX9d state is very similar in structure to the VX9a state in the near-linear regime. In fact, the vortex distributions are almost the same, except that the former state slightly breaks the four-fold symmetry. In the central five-vortex cluster, the two vortices along the $y$ axis are closer to the central vortex than the two vortices along the $x$ axis. In addition, the four edge vortices are not along the diagonals, they are instead closer to the $y$ axis. However, as the chemical potential increases the state quickly restores the four-fold symmetry around $\mu=5$, and the two ULSs therefore converge to the same nonlinear wave in the TF regime.

The VX9e state is like a distorted $3\times3$ square lattice in the near-linear regime, again with alternating charges. As the chemical potential increases, the two edge vortices are induced into the condensate at about $\mu=10$. Meanwhile, the central vortex of charge $-1$ undergoes an elongation and pair creation around $\mu=9.7$, yielding an aligned vortex triple of charges $-1, 1, -1$ along the $x$ axis and consequently a VX11 state in the TF regime. Interestingly, if we ignore the two vortices on the sides, the remaining cluster is like the VX9b state in the small-density regime.

The VX9f state is somewhat similar to the VX9b state except that the two side vortices sit at the edge of the condensate and the seven vortices are also much closer together. We briefly mention that the two edge vortices are not immediately relevant in the near-linear regime, i.e., they are not captured by the ULS. However, the two vortices emerge rather rapidly at about $\mu=4.04$ in our spatial horizon, showing that something nontrivial can happen even in the near-linear regime. Considering the very rapid vortex nucleation and also the two prominent density depletion wedges, we call this state a VX9 state for simplicity. As the chemical potential increases, the two edge vortices are induced into the condensate at about $\mu=8$, but the seven vortices in the central region are all pushed together, forming effectively a single vortex of charge $3$ at the center. Therefore, the state becomes an intriguing vortex triple state with relative charges $-1,3,-1$ in the TF regime. It is also interesting to compare with the evolution of the RVQ state. Maybe we can expect a vortex triple state with relative charges $-1, 4, -1$ in the TF regime bifurcating from the following $\mu_0=5$ degenerate set, and so on.

The fifth degenerate set has five basis states \statem{40}, \statem{31}, \statem{22}, \statem{13}, and \statem{04} with $\mu_0=5$. Their polynomial parts again upto a common factor are $(4x^4-12x^2+3)/(2\sqrt{6})$, 
$\sqrt{2/3}(2x^2-3)xy$, 
$(2x^2-1)(2y^2-1)/2$, 
$\sqrt{2/3}(2y^2-3)xy$, 
$(4y^4-12y^2+3)/(2\sqrt{6})$, respectively. It is clearly rather tedious to manipulate these polynomials theoretically, but our numerical setup proceeds in essentially the same way, a common feature of computational physics. The pattern formation is even more versatile, as the number of basis states increases. The ULSs are summarized as:
\begin{align}
    \varphi_{\mathrm{DS40}}^0 &\approx 0.9892\state{40}-0.1460\state{22}+0.0088\state{04}, \\
    \varphi_{\mathrm{DS31}}^0 &\approx 0.9896\state{31}-0.1440\state{13}, \\
    \varphi_{\mathrm{DS22}}^0 &\approx -0.1697(\state{40}+\state{04})+0.9708\state{22}, \\
    \varphi_{\mathrm{DS22b}}^0 &\approx 0.4557(\state{40}+\state{04}) \nonumber \\
    &+0.0524(\state{31}+\state{13})-0.7610\state{22}, \\
    \varphi_{\mathrm{XDS4}}^0 &=\frac{1}{\sqrt{2}}(\state{31}-\state{13}), \\
    \varphi_{\mathrm{RDSXDS2}}^0 &= \frac{1}{\sqrt{2}}(\state{31}+\state{13}), \\
    \varphi_{\mathrm{RDS2}}^0 &= \state{20}_P, \\
    \varphi_{\mathrm{Phiyy}}^0 &\approx 0.5971\state{40}+0.8016\state{22}-0.0274\state{04}, \\
    \varphi_{\mathrm{U2O2}}^0 &\approx 0.5344\state{40}-0.5442\state{22}-0.6466\state{04}, \\
    \varphi_{\mathrm{RDSVX^2}}^0 &= \state{12}_P, \\
    \varphi_{\mathrm{VX^4}}^0 &= \state{04}_P, \\
    \varphi_{\mathrm{XDS2VX4}}^0 &\approx 0.7071\state{31}+(0.0608+0.7045i)\state{13}, \\
    \varphi_{\mathrm{VX4a}}^0 &\approx 0.3392\state{40}+(0.0286-0.5013i)\state{31} \nonumber \\
    &+(-0.5149+0.0177i)\state{22} \nonumber \\
    &+(0.0629+0.4982i)\state{13} \nonumber \\
    &+(0.3384-0.0233i)\state{04}, \\
    \varphi_{\mathrm{VX4b}}^0 &\approx 0.3797\state{40}-0.5004i\state{31}-0.4592\state{22} \nonumber \\
    &+0.5004i\state{13}+0.3797\state{04}, \\
    \varphi_{\mathrm{VX6}}^0 &\approx 0.4542\state{40}+(-0.0391+0.5615i)\state{31} \nonumber \\
    &-(0.2651+0.1035i)\state{22} \nonumber \\
    &+(0.2942-0.2297i)\state{13} \nonumber \\
    &-(0.3435+0.3723i)\state{04}, \\
    \varphi_{\mathrm{VX10}}^0 &\approx 0.6522\state{40}+0.7514i\state{31}+0.0506\state{22} \nonumber \\
    &-0.0848i\state{13}-0.0091\state{04}, \\
    \varphi_{\mathrm{VX12a}}^0 &\approx 0.6502\state{40}+(0.0841+0.7520i)\state{22} \nonumber \\
    &-(0.0005+0.0679i)\state{04}, \\
    \varphi_{\mathrm{VX12b}}^0 &\approx 0.5849\state{40}-0.0076i\state{31}+0.2037\state{22} \nonumber \\
    &-0.6595i\state{13}+0.4259\state{04}, \\
    \varphi_{\mathrm{VX12c}}^0 &\approx 0.5299\state{40}+(0.5571+0.0868i)\state{22} \nonumber \\
    &+(0.1181-0.6223i)\state{04}, \\
    \varphi_{\mathrm{VX12d}}^0 &\approx 0.4801\state{40}-(0.1064-0.5641i)\state{31} \nonumber \\
    &-(0.2757-0.2130i)\state{22} \nonumber \\
    &+(0.0216+0.0092i)\state{13} \nonumber \\
    &+(0.3785+0.4180i)\state{04}, \\
    \varphi_{\mathrm{VX12e}}^0 &\approx 0.4443\state{40}+(0.1869-0.3262i)\state{31} \nonumber \\
    &-(0.3484+0.4486i)\state{22} \nonumber \\
    &-(0.3623-0.1004i)\state{13} \nonumber \\
    &-(0.1099-0.4304i)\state{04}, \\
    \varphi_{\mathrm{VX12f}}^0 &\approx 0.5895\state{40}+(0.0618-0.0395i)\state{31} \nonumber \\ &+(0.1515+0.0025i)\state{22} \nonumber \\ &+(0.4782-0.0694i)\state{13} \nonumber \\ &-(0.0682+0.6212i)\state{04}, \\
    \varphi_{\mathrm{VX12g}}^0 &\approx 0.6343\state{40}+0.3688i\state{31}-0.0062\state{22} \nonumber \\
    &+0.6793i\state{13}+0.0119\state{04}, \\
    \varphi_{\mathrm{VX13}}^0 &\approx 0.4980\state{40}-(0.0032-0.1054i)\state{31} \nonumber \\ &-(0.4927+0.0026i)\state{22} \nonumber \\ &+(0.0011-0.6992i)\state{13} \nonumber \\ &-(0.0957-0.0021i)\state{04}, \\
    \varphi_{\mathrm{VX14a}}^0 &\approx 0.5458\state{40}-(0.0972+0.2757i)\state{31} \nonumber \\ &-(0.0796+0.4763i)\state{22} \nonumber \\  &+(0.0022-0.2923i)\state{13} \nonumber \\ &-(0.5161-0.1776i)\state{04}, \\
    \varphi_{\mathrm{VX14b}}^0 &\approx 0.6242\state{40}+(0.3190+0.0757i)\state{31} \nonumber \\    &-(0.0578-0.0476i)\state{22} \nonumber \\ &-(0.0132+0.3276i)\state{13} \nonumber \\ &+(0.1196-0.6127i)\state{04}, \\
    \varphi_{\mathrm{VX14c}}^0 &\approx 0.6353\state{40}-(0.1176+0.3893i)\state{31} \nonumber \\ &-(0.1371-0.2855i)\state{22} \nonumber \\ &-(0.0196-0.3254i)\state{13} \nonumber \\ &+(0.0756-0.4676i)\state{04}, \\
    \varphi_{\mathrm{VX16a}}^0 &\approx 0.4774(\state{40}+\state{04})-0.1522\state{22} \nonumber \\ 
    &+0.5104i(\state{31}+\state{13}), \\
    \varphi_{\mathrm{VX16b}}^0 &\approx 0.5188(\state{40}+\state{04}) \nonumber \\ 
    &+(0.0903+0.6735i)\state{22}, \\
    \varphi_{\mathrm{VX16c}}^0 &\approx 0.6949\state{40}-(0.1222+0.1384i)\state{22} \nonumber \\ &+(-0.0861+0.6896i)\state{04}, \\
    \varphi_{\mathrm{VX16d}}^0 &\approx 0.4903\state{40}+(0.0613-0.5029i)\state{31} \nonumber \\ &+(0.0722-0.0264i)\state{22} \nonumber \\ &+(0.3716+0.3444i)\state{13} \nonumber \\ &+(0.3744-0.3166i)\state{04}, \\
    \varphi_{\mathrm{VX16e}}^0 &\approx 0.6528\state{40}+(0.0795-0.6507i)\state{22} \nonumber \\ &-(0.0609-0.3748i)\state{04}, \\
    \varphi_{\mathrm{VX16f}}^0 &\approx 0.6404\state{40}+(0.2573+0.0966i)\state{31} \nonumber \\ &-(0.1275-0.1119i)\state{22} \nonumber \\ &-(0.0625+0.2676i)\state{13} \nonumber \\ &+(0.0831-0.6350i)\state{04}.
\end{align}
Here, the RRS and the CRS solvers work at a slightly larger chemical potential $\mu=5.04$ because of the larger $\mu_0$. It seems likely that $\mu=5.03$ also works well. The number of independent runs shall be summarized later hereafter for simplicity.

\begin{figure*}
\includegraphics[width=\textwidth]{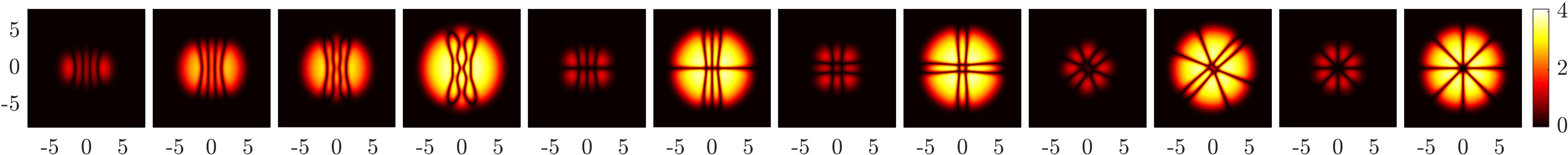}
\includegraphics[width=\textwidth]{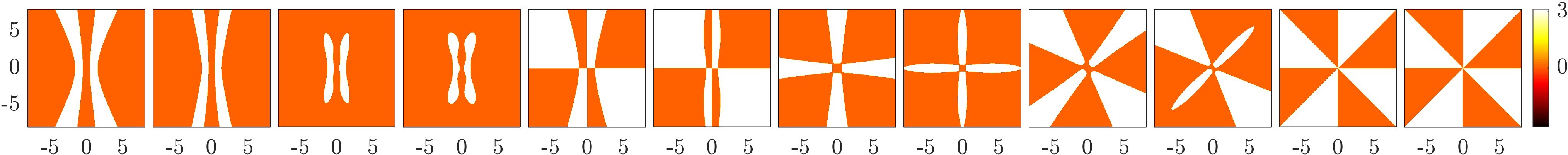}
\includegraphics[width=\textwidth]{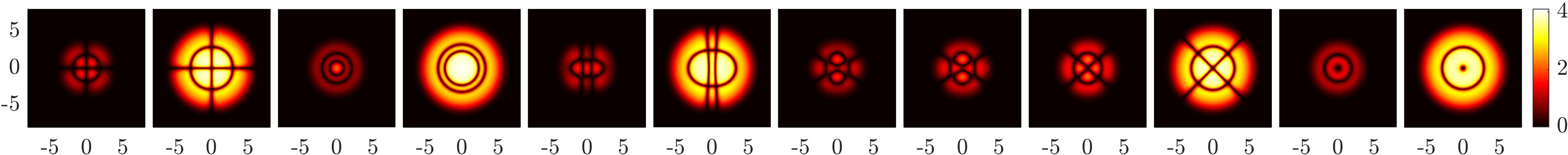}
\includegraphics[width=\textwidth]{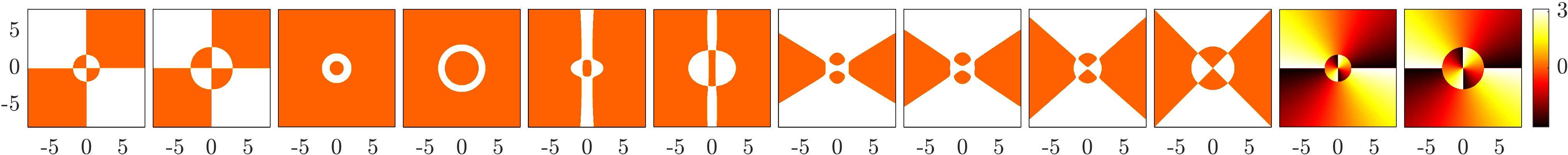}
\includegraphics[width=\textwidth]{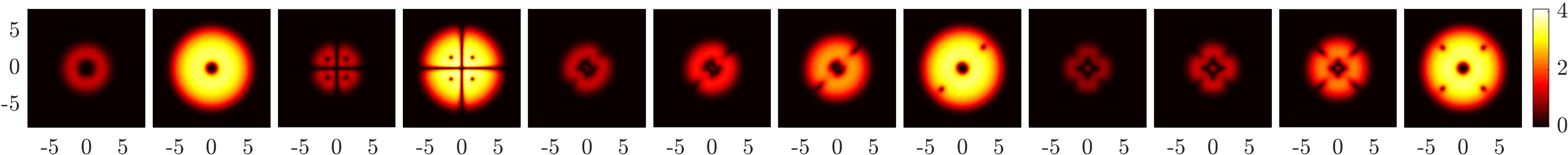}
\includegraphics[width=\textwidth]{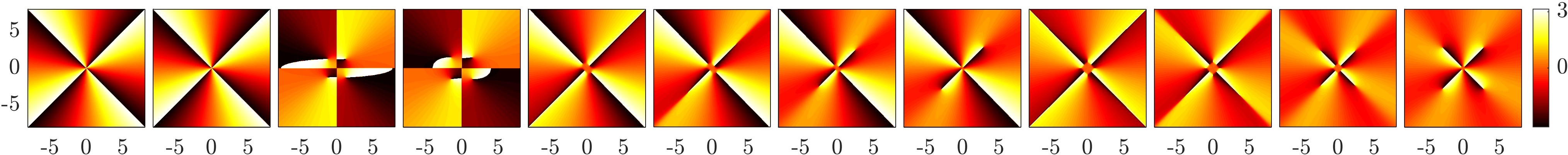}
\includegraphics[width=\textwidth]{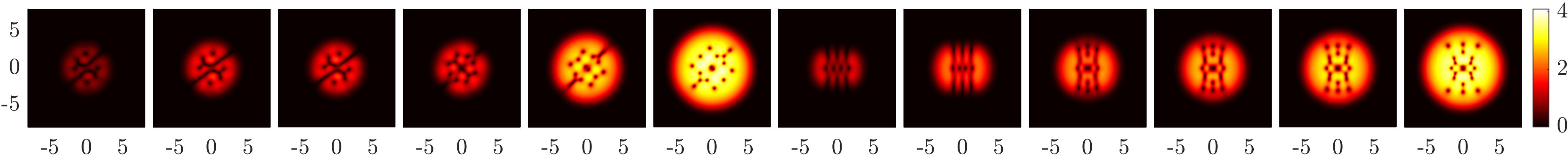}
\includegraphics[width=\textwidth]{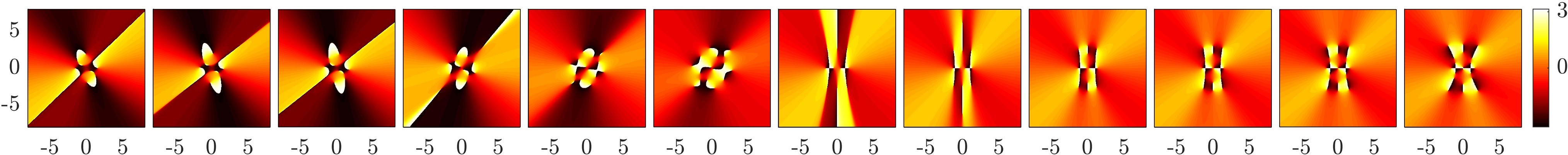}
\caption{
Solitary waves continued from the $\mu_0=5$ linear degenerate states, here typical states at $\mu=6$ and $\mu=16$ are illustrated. In the first set, the DS40 is chaotic, and states at $\mu=6, 10, 11, 16$ are shown. In the second set, the U2O2 states at $\mu=6, 6.5, 7, 16$ are shown. In the third set, the VX4a states at $\mu=6, 7, 10, 16$ and the VX4b states at $\mu=5.5, 6, 9, 16$ are depicted. In the fourth set, the VX6 states at $\mu=5.4, 6.5, 6.598, 6.600, 11.5, 16$ and the VX10 states at $\mu=6, 8, 9, 10, 12, 16$ are depicted. See also Fig.~\ref{VX12aa} and Fig.~\ref{VX14aa}.
}
\label{DS40}
\end{figure*}

It is striking that the DS40 state is already highly chaotic, and it no longer exists deep in the TF regime, at least in the isotropic trap. The structure remains qualitatively intact for a reasonably good chemical potential interval, before its wave pattern changes dramatically between $\mu=10$ and $\mu=11$. The two dark soliton stripes on both sides connect on their ends to form two closed loops at $\mu \approx 10.6$, and the two loops are subsequently induced into the condensate. The loops are highly deformed, like two peanuts facing each other at $\mu=16$. 
On the other hand, the DS31 and DS22 states appear to be more robust in their existence. Nevertheless, they may also undergo chemical potential chaos at yet larger chemical potentials according to their curved phase profiles. Similarly, even the DS20, DS30, and DS21 states may become chaotic at sufficiently large chemical potentials. This suggests that approximately parallel dark soliton stripes or planes may undergo nontrival connections in the TF regime. By contrast, the DS11 state appears to be fully robust in its existence, like other XDSs. The DS40 is stable when it is sufficiently close to the linear limit \cite{Panos:DC1}.

It is remarkable that there exists a DS22b state, it is rather similar to the DS22 state in the near-linear regime except that it slightly breaks the four-fold symmetry. Note that the central droplet of the DS22b state has approximately a rectangular shape. This state is subject to exponential instabilities in the vicinity of the linear limit \cite{Panos:DC1}. As the chemical potential grows, the difference between the two waves becomes more prominent. The two dark soliton filaments along approximately $y=x$ close on both ends at $\mu \approx 12.8$. This example illustrates that a single Cartesian basis state may correspond to more than one solitary waves in the near-linear regime. For completeness, we briefly mention that there appears to be \textit{no} solitary wave corresponding to the linear \statem{32} state in the next linear degenerate set, in agreement with \cite{Panos:DC1}.

The XDS4 and RDSXDS2 states are naturally expected from the associated polar states. It is interesting that the ULS of the RDSXDS2 is qualitatively the real or imaginary parts of the RDSVX$^2$ state, contrary to the complicated structure of the $\phi$ soliton with only a single dark soliton stripe. Similarly, the ULS of XDS4 is qualitatively the real or imaginary parts of the single vortex state of charge $4$ as mentioned earlier. Both states are subject to exponential instabilities in the vicinity of the linear limit \cite{Panos:DC1}.

The RDS2 is a polar state, and therefore it is the ULS of itself. While the RDS2 remains relatively simple in structure, its spectrum is rather complex. It is unstable right from the linear limit, but again it can be fully stabilized with suitable external potential barriers \cite{Wang:RDS2}. The double $\phi$ soliton Phiyy is quite interesting, as it is a real, hybrid Cartesian and polar state, the deformed RDS accounts for two units of energy and the two dark soliton stripes like a DS20 state account for the other two units of energy. Note that the double $\phi$ soliton is not an associated polar state, and it appears to be exponentially unstable in the vicinity of the linear limit \cite{Panos:DC1}. More such hybrid states arise at higher energies such as the triple $\phi$ state Phiyyy or the Phiyyx from the next $\mu_0=6$ degenerate set \cite{Panos:DC1}. These two states essentially have a deformed RDS in coexistence with a DS30 state and a DS21 state, respectively.

The U2O2 dark soliton state is quite intriguing with two relatively small off-center ring dark solitons next to each other and also two small U dark solitons at the sides. This state is subject to an exponential instability in the vicinity of the linear limit \cite{Panos:DC1}.
Its continuation is chaotic, the dark soliton filaments undergo complicated connections between about $\mu=6$ and $\mu=7$. The two RDSs are pushed next to each other and they also connect with the two U dark solitons, forming remarkably a RDSXDS2 structure in the TF regime. This is simply a rotated version of the RDSXDS2 state above. The following two complex waves are the RDSVX$^2$ and the VX$^4$ polar states found earlier.

The XDS2VX4 state is like a XDS2 in coexistence with a SVQ, and its chemical potential continuation is quite robust. It is similar in pattern to the DSVX4 state stemming from the $\mu_0=4$ degenerate set but with an additional dark soliton stripe. The charge of the four vortices again alternates as expected. The state appears to be quite symmetric, and the vortices are located at approximately $(\pm 1.22, \pm 1.22)$ in the near-linear regime.

The VX4a state has four vortices of charge $-1$ forming a small vortex cluster at the center of the condensate and two small density depletion regions at the edge in the small-density regime. As the chemical potential grows, two edge vortices of the opposite charge $1$ are nucleated therein at about $\mu=6.1$. These two vortices are then induced into the condensate at about $\mu=10$. Meanwhile, the four central vortices merge together, forming effectively a single vortex of charge $-4$ in the TF regime. As expected earlier, we indeed find a vortex triple state with relative charges $1, -4, 1$ following the previous pattern of the vortex triple states in the TF regime.

The VX4b state is very similar to the VX4a state, except that there are four small density depletion regions at the edge in the small-density regime. As the chemical potential increases, four edge vortices of the opposite charge $1$ are nucleated therein at $\mu \gtrsim 5.5$. These four vortices are induced into the condensate at about $\mu=9$. We mention in passing that four pair creations also occur at $\mu \approx 5.6$, see the state of $\mu=6$ along the $x$ and $y$ axes. However, they are pretty short-lived and then annihilate again at $\mu \approx 6.9$. The four central vortices are again pushed together, forming a single vortex of charge $-4$ in the TF regime. A similar state with a central vortex of charge $-3$ is obtained from the previous degenerate set, but the ULSs are very different in nature.

The VX6 state is very intriguing in its pattern formation, and the chemical potential evolution is also quite radical due to the strong density depletion regions. There are only $6$ vortices in the vicinity of the linear limit, but as the chemical potential increases a pair creation quickly occurs on each side at about $\mu=5.08$, leading to a VX10 state. For the right bottom condensate, the lower leg of the Y structure is where the pair creation occurs, and so on. It seems that the Y structure is a very common vortex cluster in the solitary wave pattern formation, it has a central vortex of charge $\pm 1$ surrounded by three approximately equally spaced vortices of charge $\mp 1$ \cite{Panos:DC1}. Next, an edge vortex along each density wedge is nucleated at about $\mu=5.5$, leading to a VX12 state. These two edge vortices first move closer to the condensate, but then they leave the condensate again at about $\mu=6.592$, and shortly after a sudden structural transition occurs between $\mu=6.598$ and $\mu=6.600$. Two vortices are induced at each density depletion wedge and the vortices reorganize strongly to maintain equilibrium, acquiring also a reflection symmetry. It should be noted that this discontinuous transition is not a genuine chemical potential evolution, but rather it is the evolution of the Newton's solver. To our knowledge, this appears to be a sudden transition rather than a gradual crossover, note the narrow chemical potential range. This suggests that the former VX10 state ceases to exist beyond a critical $\mu_c\approx6.598$ and it morphs into the latter VX14 state to maintain equilibrium. We shall discuss this VX14a state again later, which itself has a linear limit. Finally, the two edge vortices leave the condensate at about $\mu=10.4$ and a pair creation occurs at approximately $\mu=11.2$ at each density depletion edge to heal the condensate, making a total of $16$ vortices. The central two vortices of the same charge also tend to merge in the TF regime. The process after the transition is a genuine chemical potential evolution of the VX14a state. We have shown the transition and the subsequent evolution here for completeness, and the readers who are not willing to establish such a link can ignore the latter part. It is rather striking that the chemical potential evolution or the continuation of a solitary wave can be so complex and intriguing.

The VX10 state is largely from a complex mixing of the \statem{40} and \statem{31} states, explaining the four aligned vortices of the same charge $1$ along the $x$ axis and the strong density depletion regions on both sides. There are, however, other contributions such that three vortices of charge $-1$ sit on both sides in the density depletion regions. It is like an extended version of the VX7 state from the previous set in the weakly-interacting regime. The VX10 state appears to be oscillatorily unstable in the vicinity of the linear limit \cite{Panos:DC1}. As the chemical potential increases, another vortex of charge $1$ is induced along the $y$ axis on each side at about $\mu=8.0$ to heal the condensate, generating a VX12 state. Finally, four pair creations occur inside the condensate at about $\mu=10.3$, leading to a VX20 state. The two central vortices of the same charge are pushed together in the TF regime.

\begin{figure*}
\includegraphics[width=\textwidth]{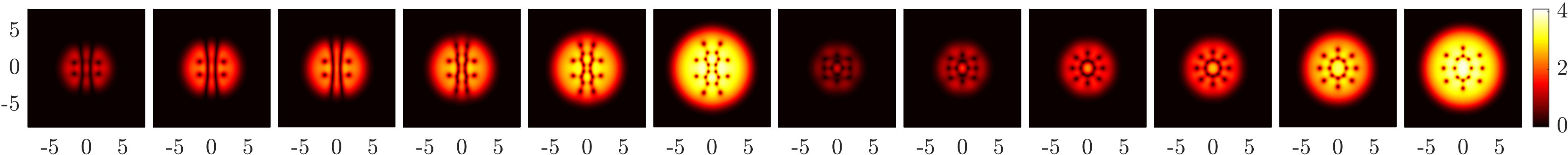}
\includegraphics[width=\textwidth]{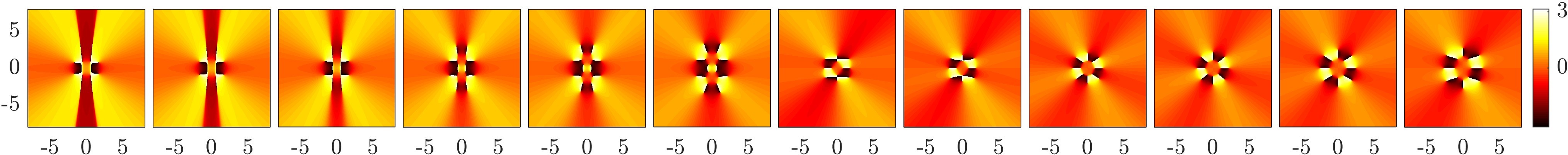}
\includegraphics[width=\textwidth]{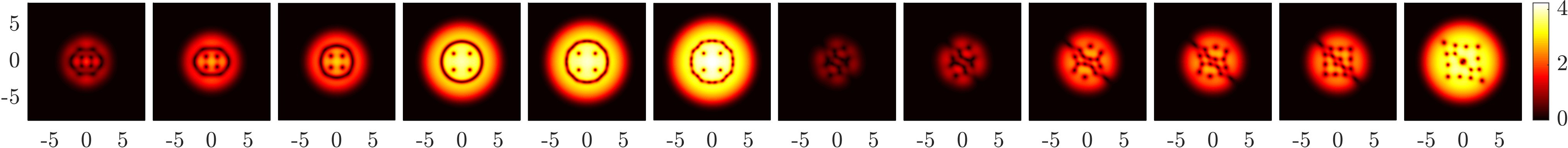}
\includegraphics[width=\textwidth]{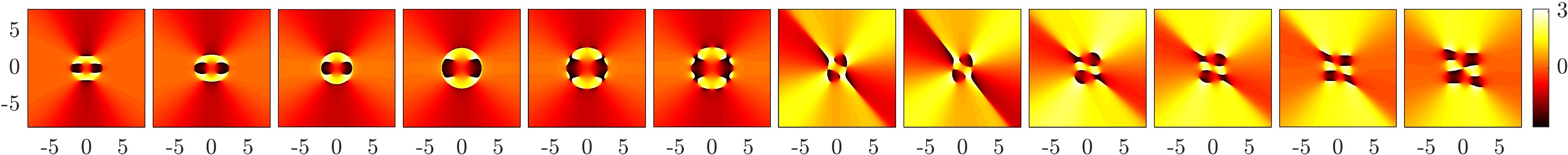}
\includegraphics[width=\textwidth]{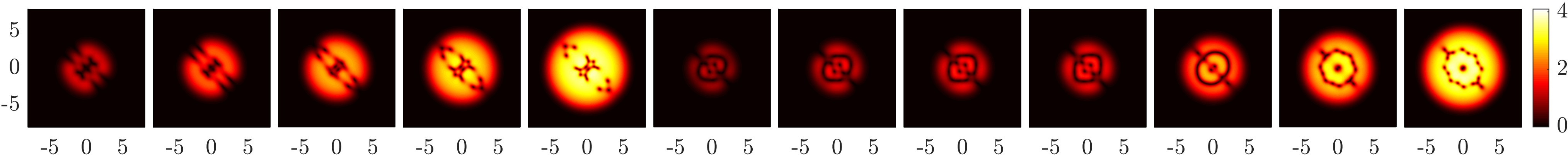}
\includegraphics[width=\textwidth]{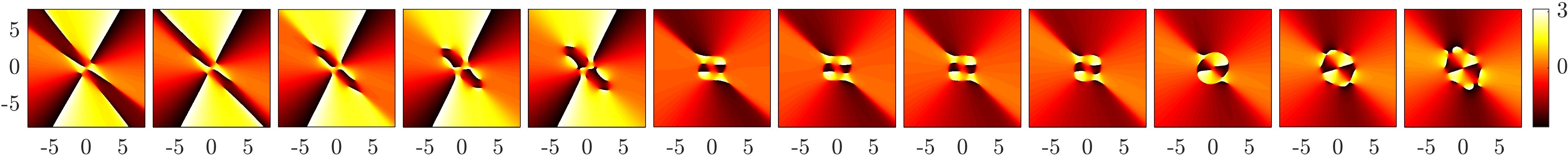}
\includegraphics[width=\textwidth]{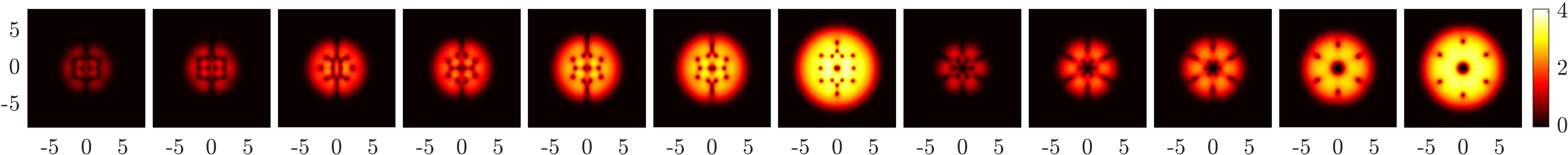}
\includegraphics[width=\textwidth]{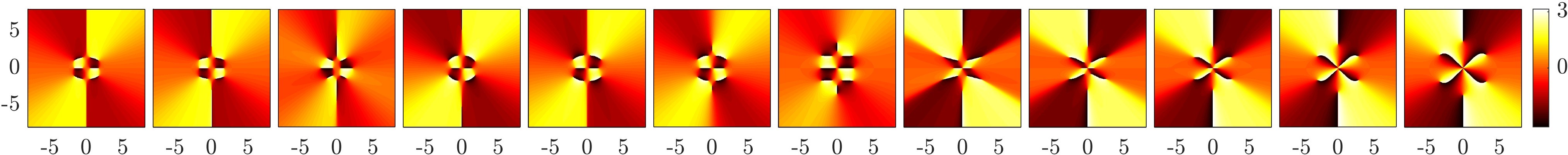}
\caption{
Solitary waves continued from the $\mu_0=5$ linear degenerate states, continued from Fig.~\ref{DS40}.
In the first set, the VX12a states at $\mu=6, 8, 9, 10, 12, 16$ and the VX12b states at $\mu=5.5, 6, 7, 8, 12, 16$ are depicted. In the second set, the VX12c states at $\mu=6, 8, 9, 15, 16, 18$ and the VX12d states at $\mu=5.4, 6, 9, 9.2, 9.3, 16$ are depicted. The third set illustrates the VX12e states at $\mu=6, 8, 10, 12, 16$ and the VX12f states at $\mu=5.5, 6, 6.3, 6.4, 9, 12, 16$. The fourth set shows the VX12g states at $\mu=5.5, 6, 7.936, 7.938, 10.5, 11.5, 16$ and the VX13 states at $\mu=6, 8, 9, 12, 16$.
}
\label{VX12aa}
\end{figure*}

The VX12a state appears to have two dark soliton stripes in the near-linear regime, but there are four vortices along each stripe instead, leading to the density depletions therein. Two of them are closer to the center, and the other two are closer to the edge. The vortex charge alternates along and between the stripes, and also around the four prominent vortices. However, each of the inner four vortices and its horizontal prominent neighbour have the same charge. As the chemical potential increases, each inner vortex undergoes an elongation and pair creation at around $\mu=9.0$ to heal the density depletion, leading to a total of $20$ vortices. As such, all the neighbouring vortices have the opposite charge in the TF regime.

The VX12b state has $4$ vortices in the inner ring and $8$ vortices in the outer ring in the near-linear regime. The vortex charge alternates along both rings, and the neighbouring vortices between the rings have the opposite charge along the $x$ axis, but the same charge along the $y$ axis. However, each inner vortex along the $y$ axis undergoes an elongation and pair creation at around $\mu=6.2$ such that all the neighbouring vortices have the opposite charge, yielding a VX$16$ state. It is like two concentric vortex necklaces, each has $8$ vortices. Interestingly, the state slightly breaks the eight-fold symmetry in the TF regime.

The VX12c state also has $4$ vortices in the inner ring and $8$ vortices in the outer ring in the near-linear regime. The vortex charge alternates along both rings, and the neighbouring vortices between the rings have the opposite charge horizontally, but the same charge vertically. As the chemical potential increases, each vortex-anti-vortex pair of the outer ring along the $x$ axis annihilates at about $\mu=8.4$, leading to a more symmetric VX8 state with strong density depletions. It is important to note that the dark ring is \textit{not} a RDS, rather it has $4$ vortices with the density depletion regions in between them. The density depletion persists over a large chemical potential interval until further vortex nucleation. Particularly, each diagonal vortex undergoes an elongation and pair creation around $\mu=15.3$ and four pair creations occur along the horizontal and vertical directions around $\mu=15.5$, yielding a VX24 state in the TF regime. We have continued this state to a slightly larger chemical potential $\mu=18$ to better illustrate the nucleated vortices.

The VX12d state has a pair of edge vortices, three pairs of the inner vortices are reasonably prominent, and each of the larger dark regions has two vortices of the same charge, which is opposite to that of the edge vortices. Then, the vortex closer to the edge vortex undergoes an elongation and pair creation at about $\mu=5.5$, yielding the familiar Y structure. As such, the state becomes a VX16 state. As the chemical potential increases, the state starts to evolve dramatically from about $\mu=9$ until $\mu\lesssim 9.3$, the vortices gradually organize into a configuration with pattern as depicted at $\mu=9.3$. The two edge vortices are induced into the condensate at about $\mu=10$. Interestingly, the state does not evolve into the $4 \times 4$ square lattice, but rather the two central vortices are pushed closer together, forming effectively a single vortex of charge $2$ in the TF regime.

\begin{figure*}
\includegraphics[width=\textwidth]{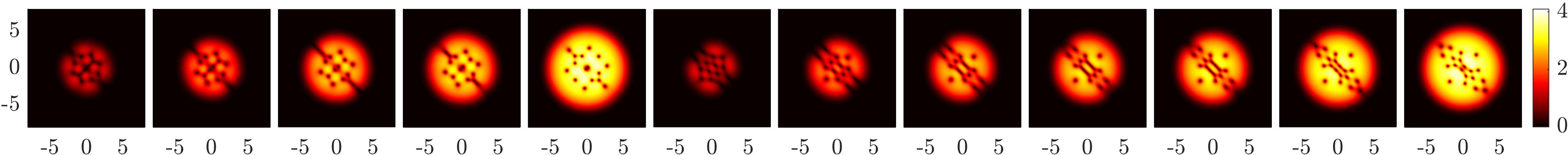}
\includegraphics[width=\textwidth]{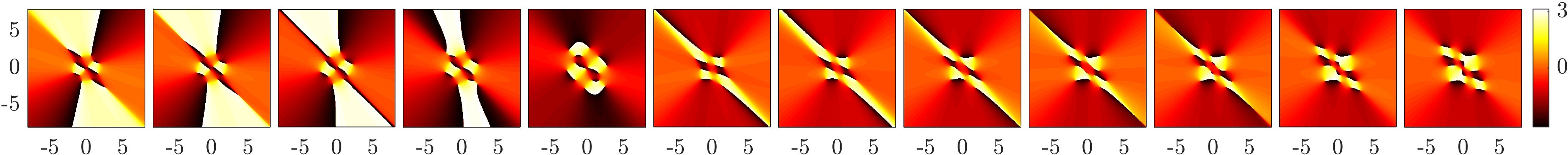}
\includegraphics[width=\textwidth]{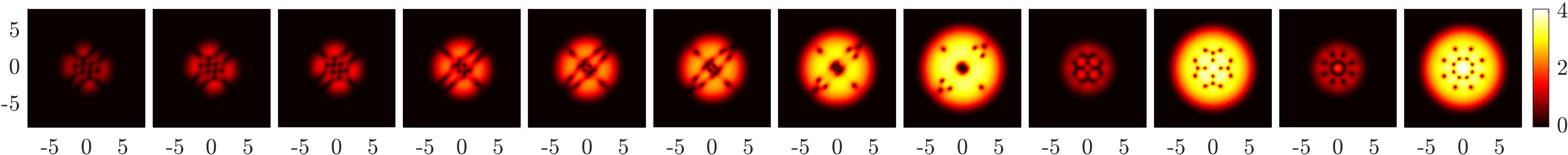}
\includegraphics[width=\textwidth]{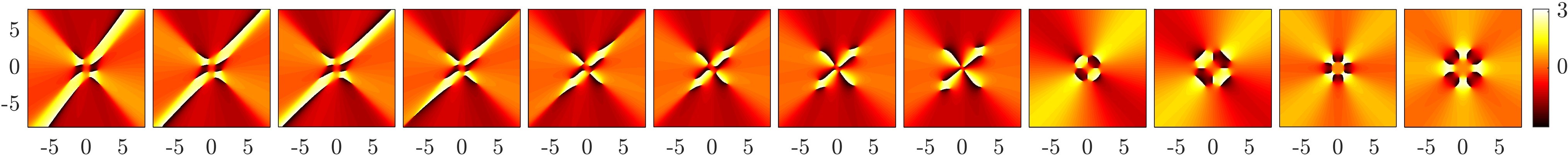}
\includegraphics[width=\textwidth]{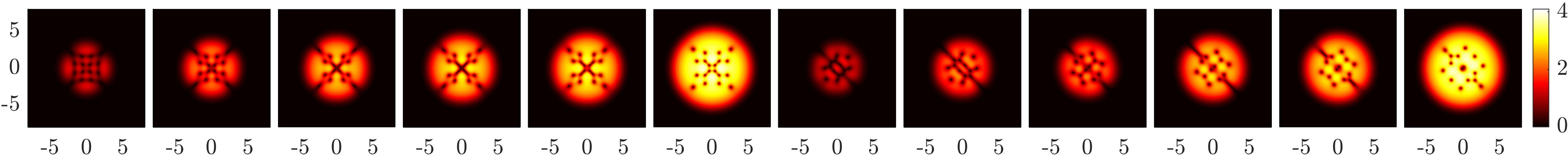}
\includegraphics[width=\textwidth]{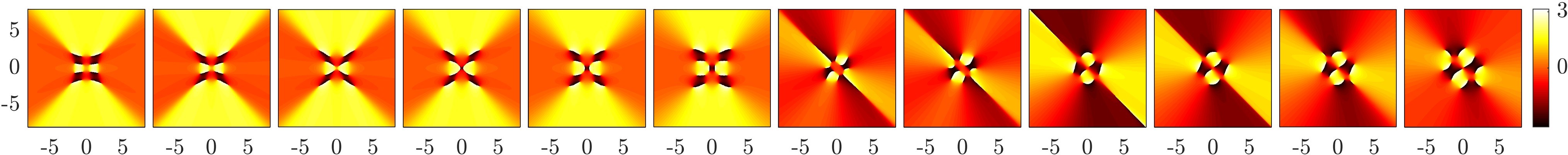}
\includegraphics[width=\textwidth]{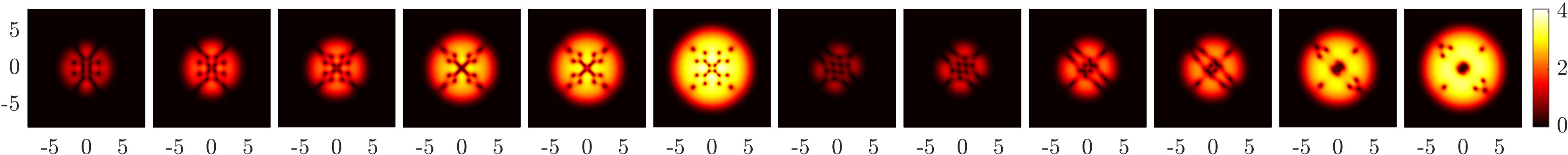}
\includegraphics[width=\textwidth]{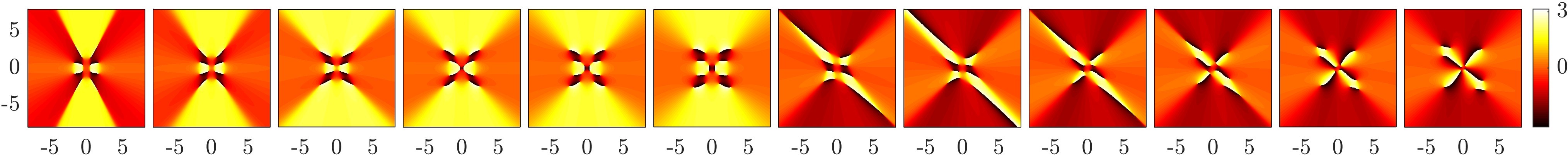}
\caption{
Solitary waves continued from the $\mu_0=5$ linear degenerate states, continued from Fig.~\ref{DS40} and Fig.~\ref{VX12aa}.
The first set shows the VX14a states at $\mu=6, 8, 10.5, 11.5, 16$ and the VX14b states at $\mu=6, 8, 10, 11, 12, 14, 16$. The second set depicts the VX14c states at $\mu=5.5, 6, 6.2, 8, 9, 10, 13, 16$, the VX16a states at $\mu=6, 16$, and the VX16b states at $\mu=6, 16$. The third set illustrates the VX16c states at $\mu=6, 8, 10, 11, 12, 16$, and the VX16d states at $\mu=6, 7.934, 7.936, 10.5, 11.5, 16$. The fourth set shows the VX16e states at $\mu=6, 7.5, 8, 11, 12, 16$, and the VX16f states at $\mu=5.5, 6, 8, 9, 13, 16$.
}
\label{VX14aa}
\end{figure*}

The VX12e state has eight vortices in the central region and four edge vortices of charge $1$ in the near-linear regime. The two vortices in the center have charge $-1$ and the three vortices on each side have charges $-1, 1, -1$, respectively. Both three vortex clusters merge together at $\mu \approx 8$, yielding a single vortex of charge $-1$ on each side. As such, the system turns into a VX8 state. Next, an additional vortex of charge $-1$ is nucleated at $\mu \approx 8.7$ in our spatial horizon in each density depletion region, leading to a VX10 state. The four former edge vortices and the two latter edge vortices are induced into the condensate at about $\mu=10, 12$, respectively. In addition, four pair creations occur near the two merged vortices above at $\mu\approx 12.0$, yielding finally a VX18 state in the TF regime. While the evolution is somewhat complicated, the continuation is nevertheless robust and smooth.

The VX12f state is quite complicated in the near-linear regime, it has two inner vortices near the center, two edge vortices, and four vortices along each of the two density depletion arms. As the chemical potential increases, two pair creations occur at $\mu \approx 6.1$ at the arms along approximately the $y=1.8x$ line, yielding a VX16 state. In addition, the state quickly takes a more symmetric configuration from $\mu \approx 6.4$. Interestingly, the state then appears to have a deformed RDS. However, it is a connected density depletion region with vortices, there is no genuine RDS there. The two edge vortices are induced into the condensate at $\mu \approx 9$. Then, two elongation and pair creations occur at $\mu \approx 11.2$ at the ring along the $y=-x$ line, creating two Y solitons therein and consequently a VX20 state. Finally, the two inner vortices are pushed together, forming effectively a single vortex of charge $-2$ in the TF regime. 

The VX12g state is like a deformed $3 \times 4$ vortex lattice with alternating charges except crossing the $y$ axis in the near-linear regime. As the chemical potential increases, two edge vortices are nucleated at $\mu \approx 5.6$, leading to a VX14 state. This structure then deforms and two more edge vortices are nucleated at $\mu \approx 7.8$, yielding a VX16 state. Then, the state becomes progressively similar to the VX16d state until it makes a discontinuous state transition into the VX14a state between $\mu=7.936$ and $\mu=7.938$. The VX16d state has its own linear limit, and we shall discuss it later. The VX14a state was also encountered earlier in the chaotic evolution of the VX6 state. The subsequent evolution is therefore the same, i.e., the two edge vortices leave the condensate at about $\mu=10.4$ and a pair creation occurs at approximately $\mu=11.2$ at each density depletion edge, yielding a total of $16$ vortices in the TF regime.

The VX13 state has a central vortex of charge $-2$ surrounded by a ring of $6$ vortices of charge $1$ and then another ring of $6$ vortices of charge $-1$ in the near-linear regime. The two necklaces of vortices are relatively rotated by $\pi/6$. As the chemical potential increases, the inner seven vortices are gradually pushed together, forming effectively a single vortex of charge $4$ in the TF regime. The $6$ edge vortices are induced into the condensate at about $\mu=9$.

The VX14a state appears to be a highly robust structure in existence, as it was already encountered twice as a final product of the chaotic evolution of the VX6 and the VX12g states, indeed see also the chaotic evolution of the VX16d state. Therefore, its continuation is largely known at this stage, i.e., the two edge vortices leave the condensate at $\mu \approx 10.4$ followed by two pair creations therein at $\mu \approx 11.2$, yielding finally a VX16 state in the TF regime.

The VX14b state is like a miniature vortex lattice in the near-linear regime, there are $2, 3, 4, 3, 2$ vortices, respectively, in the successive $y=x$ lattice planes. The vortices are of the same charge in a lattice plane and the vortex charge alternates between the neighbouring lattice planes. As the chemical potential increases, the four edge vortices are induced into the condensate at about $\mu=10$, but soon two further edge vortices are nucleated at $\mu \approx 10.6$ to heal the edge density depletion, leading to a VX16 state. Next, each of the two central vortices undergoes an elongation and pair creation at $\mu \approx 10.8$, yielding a VX20 state. The two edge vortices are also induced into the condensate at about $\mu=14$.

The VX14c state has $6$ edge vortices of charge $1$, and the inner $3, 2, 3$ vortices along approximately the $y=-x$ direction in turn have charges $-1, 1, -1$ in the near-linear regime. As the chemical potential increases, the vortices adjust their positions slightly and the state takes a more symmetric form at $\mu \approx 6.1$. Two more vortices are nucleated in our spatial horizon at $\mu \approx 7.8$, yielding a VX16 state, and the former $6$ edge vortices are induced into the condensate at about $\mu=9$. Soon, two pair annihilations occur inside the condensate at $\mu \approx 9.1$, leading to a VX12 state. The latter $2$ edge vortices are induced into the condensate at about $\mu=12.5$, and the $4$ central vortices merge together, forming effectively a single vortex of charge $-4$ in the TF regime.

The chemical potential continuation of the VX16a and VX16b states is quite robust. The VX16a state has four Y solitons, and the central vortex charge alternates between the neighbouring Y solitons. The VX16b state has two concentric vortex necklaces, and each ring has eight vortices. The vortex charge alternates along each ring, and also between the neighbouring vortices of the two rings. Both states are subject to oscillatory instabilities in the vicinity of the linear limit \cite{Panos:DC1}.

The VX16c state is like a deformed square lattice of alternating charge between the neighbouring vortices in the small-density regime. As the chemical potential increases, the structure deforms further. The $4$ edge vortices are induced into the condensate at about $\mu=10$. Next, the four central vortices come together and annihilate at $\mu \approx 10.8$, leading to a VX12 state with a central density depletion region. Four pair creations quickly follow therein at $\mu \approx 11.1$, yielding a VX20 state in the TF regime.

The VX16d state has $2$ edge vortices of charge $1$, and $5, 4, 5$ inner vortices along approximately the $y=x$ direction in the near-linear regime. The vortex charge alternates along the outer shell of the $8$ vortices and also along the three aligned vortex sets above, except that the two central vortices have the same charge of $-1$. We have actually encountered this state previously, i.e., the VX12g state gradually morphs into the VX16d state. As such, it makes essentially the same discontinuous transition into the VX14a state between $\mu=7.934$ and $\mu=7.936$, the rather minor difference of the critical chemical potential is likely because the states before the transition are highly similar but not exactly identical. The subsequent chemical potential evolution after the transition is the same, to our knowledge. The two edge vortices first leave the condensate at about $\mu=10.4$ and then two pair creations occur therein at approximately $\mu=11.2$, yielding a VX16 state in the TF regime. 

The VX16e state is similar in structure to the VX16c state but it is more deformed in the small-density regime. In fact, the vortices gradually adjust their equilibrium positions as the chemical potential increases, and the system takes a more symmetric form and turns into the VX16c state at $\mu \lesssim 7.8$. The subsequent evolution is therefore the same, i.e., the four edge vortices are induced into the condensate at about $\mu=10$, then the four central vortices come together and annihilate at $\mu \approx 10.8$, followed by another four pair creations therein at $\mu \approx 11.1$, yielding a VX20 state in the TF regime.

The VX16f state is very similar to the VX14c state after its symmetric transition but with two far edge vortices in the near-linear regime. They also differ in the orientation and by a charge conjugation, however, these are of litter significance. Interestingly, these two vortices are ejected out of the condensate quickly at $\mu \approx 5.6$. Therefore, the VX14c state actually converges to the VX16f state in its symmetric transition. The subsequent evolution is hence known, the $2$ vortices are again nucleated at $\mu \approx 7.8$, and the $6$ near edge vortices are induced into the condensate at about $\mu=9$. Soon, two pair annihilations occur inside the condensate at $\mu \approx 9.1$, and the $2$ edge vortices are induced into the condensate at about $\mu=12.5$, and the $4$ central vortices merge together, forming effectively a single vortex of charge $4$ in the TF regime.

The foregoing examples clearly demonstrate the power of the linear limit continuation method, finding numerous known as well as novel wave patterns such as the XDS2VX4, VX4b, VX6, VX12b, VX12c, VX13, VX14a, VX16d states to name a few, to our knowledge. Particularly, we have restricted our focus on relatively low-lying states upto \textit{only} $\mu_0=5$. The next degenerate sets of $\mu_0=6$ and $\mu_0=7$ may likely offer many more states of increasing complexity. In fact, a detailed study of these states is rather elaborate, and we shall discuss them in a separate work later. To focus on the principles of the method, we turn to anisotropic traps in the next. Again, there are too many possibilities, as $\kappa=p/q$ can be \textit{any} rational number. Therefore, we present some typical examples to demonstrate how the method works and establish the paradigm. A more systematic exploration of the program can then be carried out in future works.

\subsection{States from anisotropic traps}
\label{SAT}

To illustrate the method in a generic setting, we now systematically investigate the low-lying linear degenerate sets of the prototypical $\kappa=1/2$ lattice planes as shown in Fig.~\ref{XY}. While the rotational symmetry is broken, the system still has the reflection symmetries about the $x$ and $y$ axes, and we identify and continue distinct states upto the reflection symmetries. An additional feature here is that after the continuation of a state in the chemical potential, we also continue it in $\kappa$ into the isotropic trap in the TF regime if possible. First, the isotropic trap is rather common. Many solitary waves are studied in the isotropic trap but their linear limits are not necessarily realized in this trap, e.g., the vortex dipole state \cite{PK:DSVX,Anderson:VX2}. Second, a state may evolve nontrivially in the continuation of $\kappa$, just as in the continuation of the chemical potential. As such, this continuation helps to identify the parametric connectivity of the pertinent states in the generalized parameter space of $(\mu,\omega_x)$. To our knowledge, this is a nontrivial task, as the trap continuation is equally chaotic. Indeed, it is frequently difficult to predict a priori how a state evolves when $\kappa$ is tuned, if at all possible. We adopt a piecewise constant $\delta \omega_x$ continuation schedule for simplicity, and a typical continuation step size is $\delta \omega_x=0.005$ or smaller.



\begin{figure*}
\includegraphics[width=\textwidth]{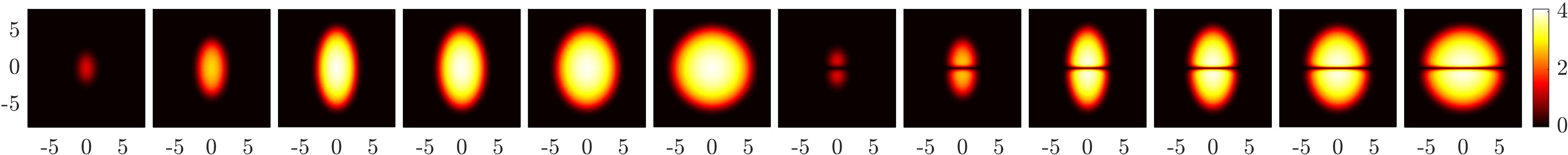}
\includegraphics[width=\textwidth]{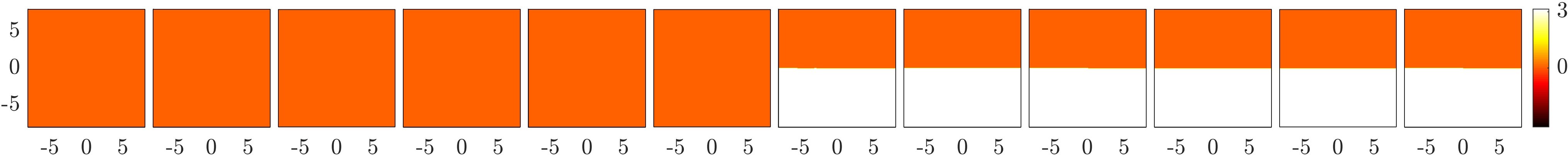}
\includegraphics[width=\textwidth]{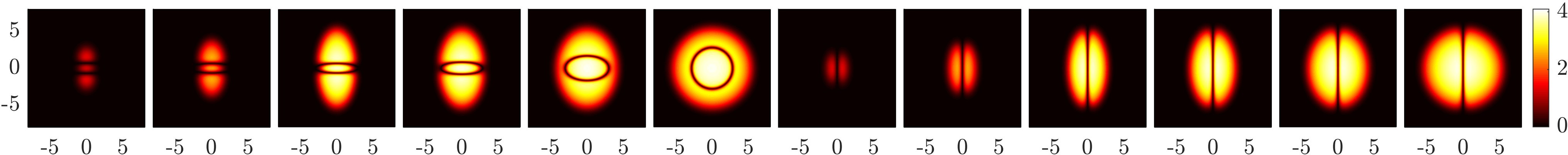}
\includegraphics[width=\textwidth]{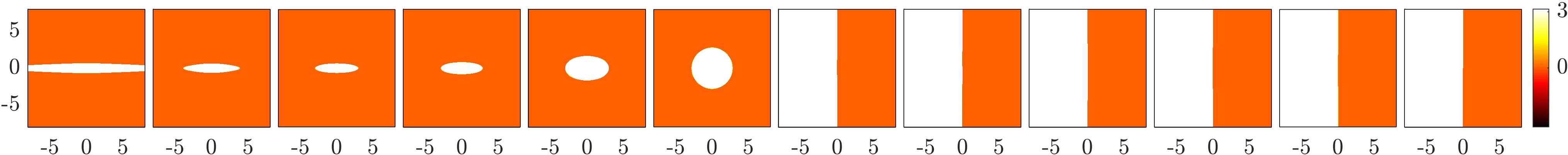}
\includegraphics[width=\textwidth]{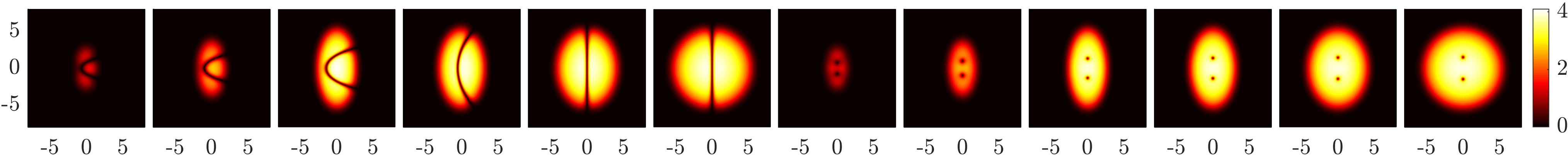}
\includegraphics[width=\textwidth]{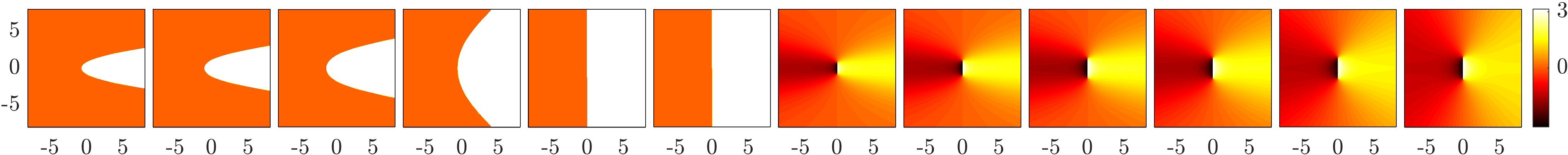}
\includegraphics[width=\textwidth]{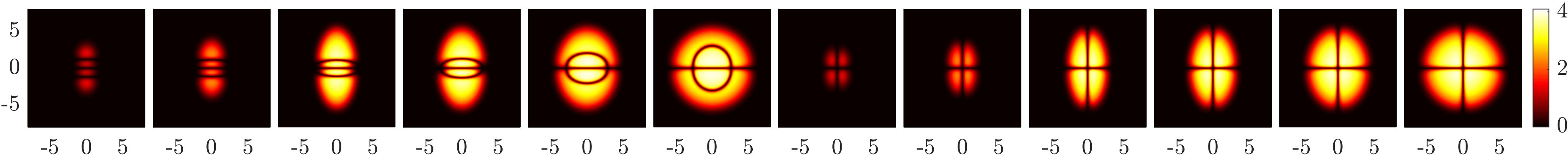}
\includegraphics[width=\textwidth]{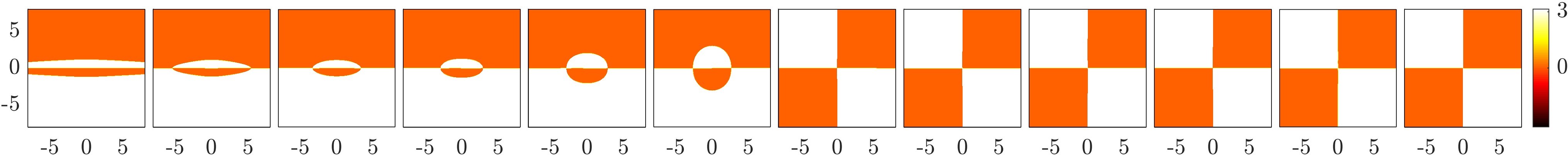}
\includegraphics[width=\textwidth]{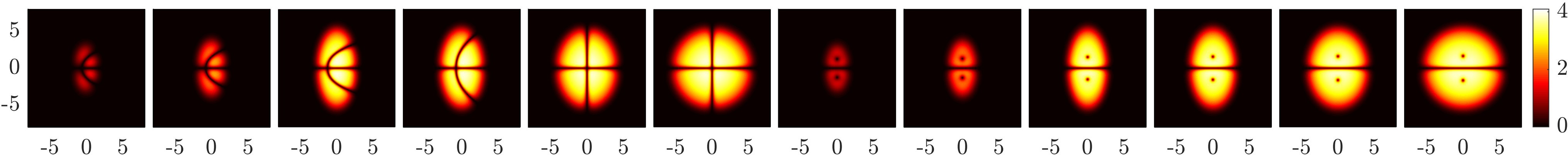}
\includegraphics[width=\textwidth]{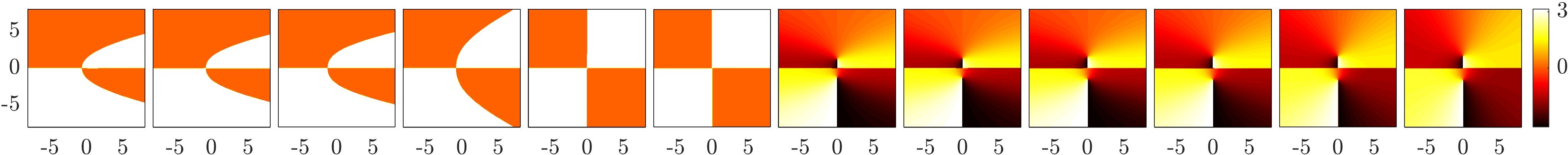}
\caption{
Solitary waves continued from the $\kappa=1/2$ linear degenerate sets. For the ground state, typical states at $\mu=2.5, 8, 16$ are shown at $\omega_x=2$, followed by a further continuation in $\kappa$ at fixed $\mu=16$. Here, the states at $\omega_x=1.7, 1.3, 1$ are depicted. The same is shown for the DS01 state, except that its first state is at $\mu=3.5$ instead.
In the second and third sets, the states are at $\mu=4.5, 8, 16$ followed by a similar continuation in $\kappa$. In the fourth and fifth sets, the chemical potentials are $\mu=5.5, 8, 16$. The continuation in $\kappa$ is again similar, except that a $\Psi$ state is shown at $\omega_x=1.8$ instead of $1.7$ for clarity.
}
\label{GSAa}
\end{figure*}

The program essentially proceeds in the same way, demonstrating the simplicity of the method. The linear excitation spectrum is $\epsilon_{mn}=(m+1/2)/\kappa+(n+1/2)=2m+n+3/2$, where $\kappa=1/2$ and $m,n=0, 1, 2, ...$ The ground state is again the state \statem{00} with $\mu_0=1.5$ and it is nondegenerate. The first excited state is the state \statem{01} with $\mu_0=2.5$, which is also nondegenerate in this particular case. It is expected that these states are parametrically connected to the ground state and the DS01 state, respectively, of the isotropic trap. Here, the DS01 state is a rotation of the DS10 state by $\pi/2$. This is numerically confirmed, as illustrated in the top set of Fig.~\ref{GSAa}. 
Naturally, the ULSs of the ground state and the DS01 state are, respectively,
\begin{align}
    \varphi_{\mathrm{GS}}^0 &= \state{00}, \quad \kappa=1/2, \\
    \varphi_{\mathrm{DS01}}^0 &= \state{01}, \quad \kappa=1/2.
\end{align}


The third linear degenerate set has two states \statem{02} and \statem{10} with $\mu_0=3.5$. The identified solitary waves in both their natural trap and the isotropic trap are shown in the second and third sets of Fig.~\ref{GSAa}. Their respective ULSs read:
\begin{align}
    \varphi_{\mathrm{DS02}}^0 &= \state{02}, \\
    \varphi_{\mathrm{DS10}}^0 &= \state{10}, \\
    \varphi_{\mathrm{U}}^0 &= 0.8375\state{02}-0.5464\state{10}, \\
    \varphi_{\mathrm{VX2}}^0 &= 0.7430\state{02}+0.6693i\state{10}.
\end{align}

Interestingly, the two dark soliton stripes of the DS02 state tend to curve towards each other rather than towards the edge of the condensate as the chemical potential grows. This is directly contrary to the DS20 state in the isotropic trap in the near-linear regime.
As such, they quickly connect into a single closed loop in the chemical potential evolution; cf. the states at $\mu=4.5$ and $\mu=8$. One remarkable consequence is that when continuing the state to $\kappa=1$ in the TF regime, it transforms into a perfect RDS state rather than the DS02 state. It is even more striking that this also happens at quite low chemical potentials at $\mu=4, 3.8$, and $3.6$, and the two dark soliton stripes consistently connect to form a perfectly circular RDS as the state is continued to $\kappa=1$. This robust transformation suggests that the DS02 state of $\kappa=1/2$ is parametrically connected to the RDS rather than the DS02 counterpart of $\kappa=1$. In this process, a series of elliptical RDSs is obtained.

The U soliton state is quite interesting, because it is a parity symmetry-breaking state sitting on one side of the condensate. This feature is absent in the states obtained in the isotropic trap. This is because in this degenerate set, \statem{02} has an even parity but \statem{10} has an odd parity. By contrast, the linear degenerate states of $\kappa=1$ always have the same parity. In the continuation to $\kappa=1$, the U soliton gradually opens a wider angle and finally becomes the DS10 state from about $\omega_x=1.66$. This shows that a solitary wave found in one particular trap may not necessarily exist in another trap.

The vortex dipole is significant here because this is the first state we have continued into the isotropic trap in which it has \textit{no} linear limit. 
In other words, a solitary wave that has no linear limit in a trap may have one in a suitable trap to apply the linear limit continuation method, and a further continuation between the traps may be available. In fact, the continuation of the vortex dipole in both the chemical potential and the trap appears to be fully robust. Because the vortex dipole has no linear limit in the isotropic trap, it should consequently have a nontrivial lower critical chemical potential below which it ceases to exist, i.e., the vortex dipole terminates before its norm reaches $0$ in the isotropic trap. Numerical simulation confirms this, finding a lower critical chemical potential $\mu_c \approx 3.47$, in line with \cite{PK:DSVX}. 
The vortex dipole is quite robust in the isotropic trap, except for a narrow interval of oscillatory instability in the chemical potential \cite{PK:DSVX}.

The fourth linear degenerate set also has two states \statem{03} and \statem{11} with $\mu_0=4.5$. The pattern formation is largely similar to that of the previous set, except there is an additional dark soliton stripe at $y=0$, as depicted in the fourth and fifth sets of Fig.~\ref{GSAa}. Their respective ULSs are summarized here:
\begin{align}
    \varphi_{\mathrm{DS03}}^0 &= \state{03}, \\
    \varphi_{\mathrm{DS11}}^0 &= \state{11}, \\
    \varphi_{\Psi}^0 &= 0.6772\state{03}-0.7358\state{11}, \\
    \varphi_{\mathrm{DSVX2}}^0 &= 0.7028\state{03}+0.7114i\state{11}.
\end{align}
Therefore, the DS03 state evolves towards the $\phi$ soliton in the TF regime. This is more evident when it is continued to $\kappa=1$. The DS11 state robustly maintains its perpendicular nature throughout the continuations. The $\Psi$ soliton is also a parity symmetry-breaking state. When it is continued to $\kappa=1$, the U soliton part again gradually opens up and crosses over to a dark soliton stripe. Therefore, the $\Psi$ soliton finally becomes the DS11 state from about $\omega_x=1.70$. The DSVX2 state is an interesting coexistence state of a vortex dipole and a horizontal dark soliton stripe. 

\begin{figure*}
\includegraphics[width=\textwidth]{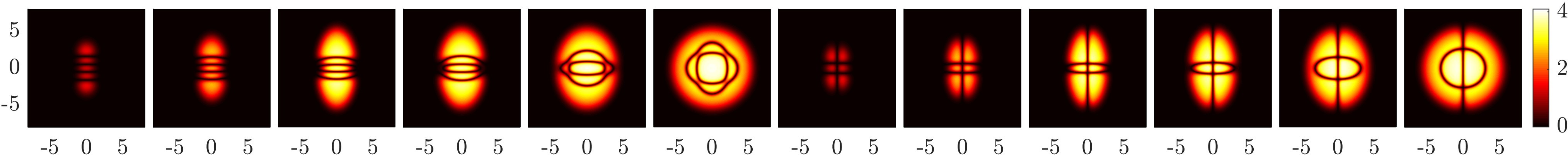}
\includegraphics[width=\textwidth]{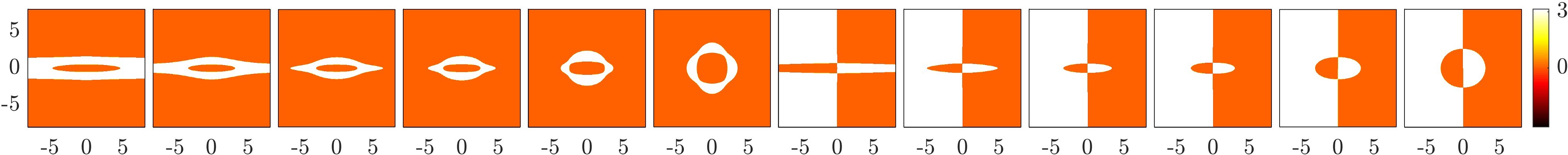}
\includegraphics[width=\textwidth]{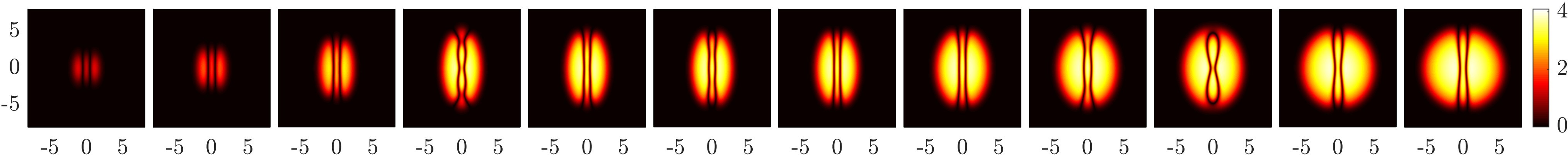}
\includegraphics[width=\textwidth]{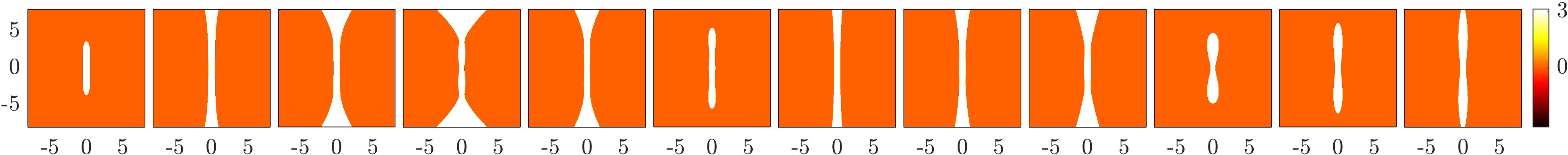}
\includegraphics[width=\textwidth]{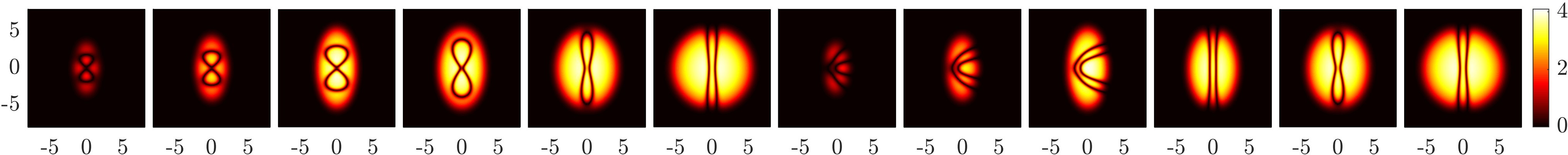}
\includegraphics[width=\textwidth]{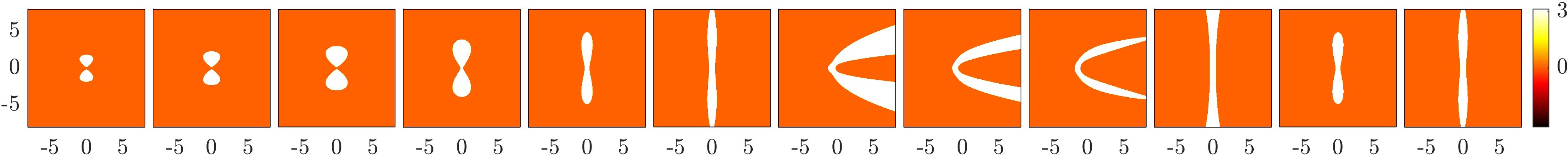}
\includegraphics[width=\textwidth]{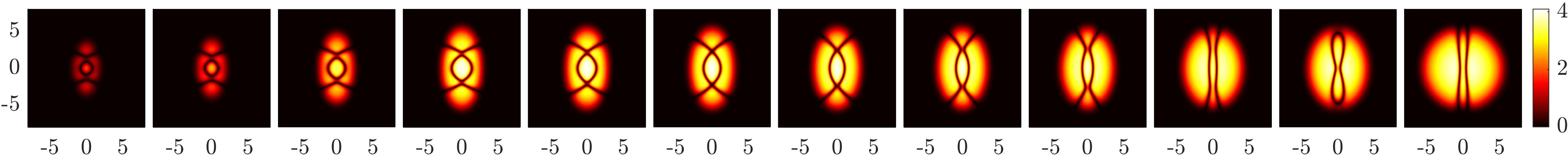}
\includegraphics[width=\textwidth]{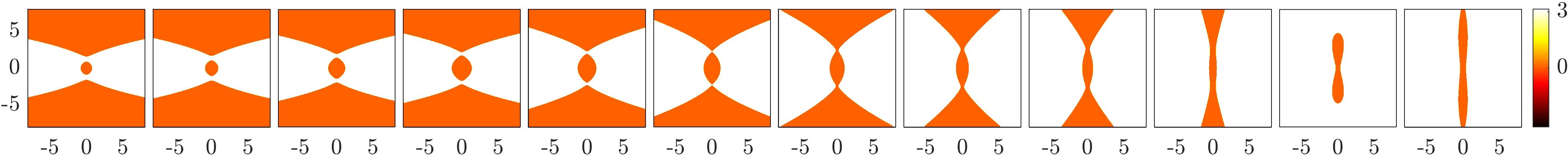}
\caption{
Solitary waves continued from the $\kappa=1/2$ linear degenerate set of $\mu_0=5.5$. For the DS04 state, typical states at $\mu=6.5, 10, 16$ are shown at $\omega_x=2$, followed by a further continuation in $\kappa$ at fixed $\mu=16$. Here, the states at $\omega_x=1.7, 1.3, 1$ are depicted. The same is shown for the DS12 state. The second set shows the DS20 state at $\mu=6.5, 8, 12, 16$, and $\omega_x=1.9150, 1.9140, 1.88, 1.5, 1.3350, 1.3340, 1.1, 1$. The third set depicts the DS8 state at $\mu=6.5, 10, 16$ and $\omega_x=1.7, 1.3, 1$, and similarly for the U2 state, except that $\omega_x=1.6$ instead of $1.7$. The fourth set illustrates the U2O state at $\mu=6.5, 8, 12, 16$, and $\omega_x=1.9, 1.8, 1.7, 1.6, 1.5, 1.3350, 1.3340, 1$. See also Fig.~\ref{VX6Aa}.
}
\label{DS04Aa}
\end{figure*}

The fifth degenerate set includes three basis states \statem{04}, \statem{12}, \statem{20} with $\mu_0=5.5$. As the degeneracy grows, the pattern formation again becomes substantially richer. 
The ULSs are summarized here:
\begin{align}
    \varphi_{\mathrm{DS04}}^0 &= 0.99997\state{04}-0.005743\state{20}, \\
    \varphi_{\mathrm{DS12}}^0 &= \state{12}, \\
    \varphi_{\mathrm{DS20}}^0 &= 0.0200\state{04}+0.9998\state{20}, \\
    \varphi_{\mathrm{DS8}}^0 &= 0.7839\state{04}+0.6208\state{20}, \\
    \varphi_{\mathrm{U2}}^0 &= 0.5254\state{04}-0.8098\state{12}+0.2609\state{20}, \\
    \varphi_{\mathrm{U2O}}^0 &= 0.7643\state{04}-0.6449\state{20}, \\
    \varphi_{\mathrm{VX6}}^0 &= 0.0405\state{04}+0.7722i\state{12}+0.6340\state{20}, \\
    \varphi_{\mathrm{VX8a}}^0 &= 0.7460\state{04}-(0.0143+0.6658i)\state{20}, \\
    \varphi_{\mathrm{VX8b}}^0 &= 0.5787\state{04}+0.6821i\state{12}+0.4470\state{20}, \\
    \varphi_{\mathrm{VX8c}}^0 &= 0.5603\state{04}+0.6792i\state{12}-0.4741\state{20}, \\
    \varphi_{\mathrm{VX8d}}^0 &= 0.6839\state{04}-0.7289i\state{12}+0.0317\state{20}, \\
    \varphi_{\mathrm{VX8e}}^0 &= 0.6974\state{04}+(0.0344+0.5691i)\state{12} \nonumber \\ 
    &-(0.0033+0.4342i)\state{20}.
\end{align}

The continuation of the DS04 state in both the chemical potential and the trap frequency is quite smooth, despite that the dark soliton filaments significantly deform in the process, as depicted in Fig.~\ref{DS04Aa}. In the near-linear regime, the inner dark soliton pair already forms a closed loop. As the chemical potential increases, the outer dark soliton pair also gradually connects into a closed loop at $\mu \approx 13.5$ in our spatial horizon, leading to a state with two highly deformed loops in the TF regime. As it is continued into the isotropic trap, it progressively morphs into a deformed RDS2 state. Strikingly, it evolves into neither the RDS2 polar state nor the two peanut state of Fig.~\ref{DS40}. This deformed RDS2 state is an isomer of the RDS2 polar state, it has two rounded square dark soliton loops, the inner one is rotated with respect to the outer one by $\pi/4$. Interestingly, this structure is similar to the RDS2 part of the associated polar states RDS2XDS and RDS2XDS2 in Fig.~\ref{polarstates2}. This state suggests that concentric dark soliton loops can form different wave patterns.

\begin{figure*}
\includegraphics[width=\textwidth]{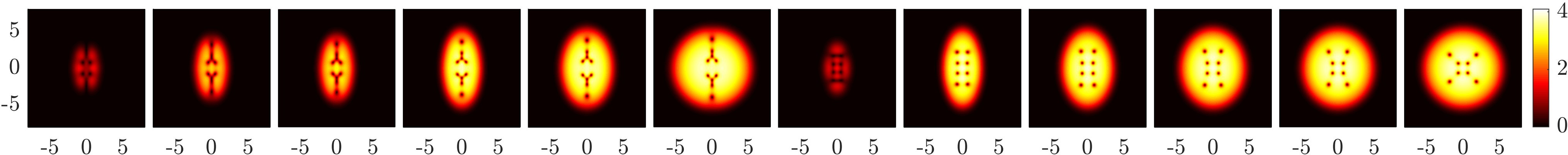}
\includegraphics[width=\textwidth]{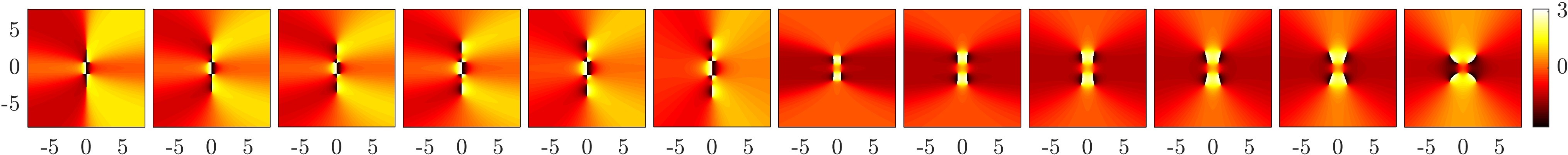}
\includegraphics[width=\textwidth]{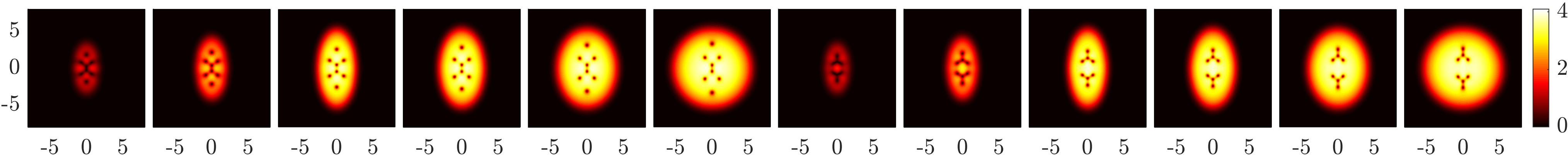}
\includegraphics[width=\textwidth]{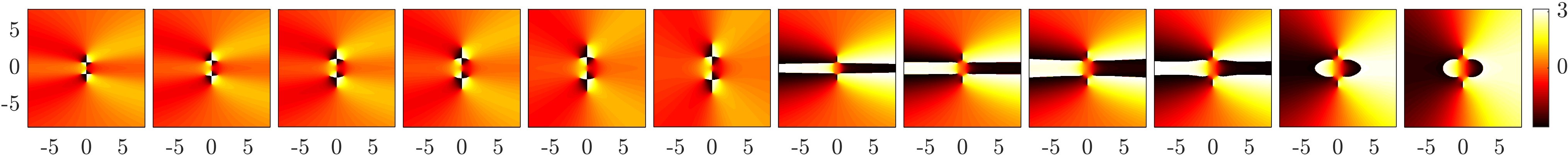}
\includegraphics[width=\textwidth]{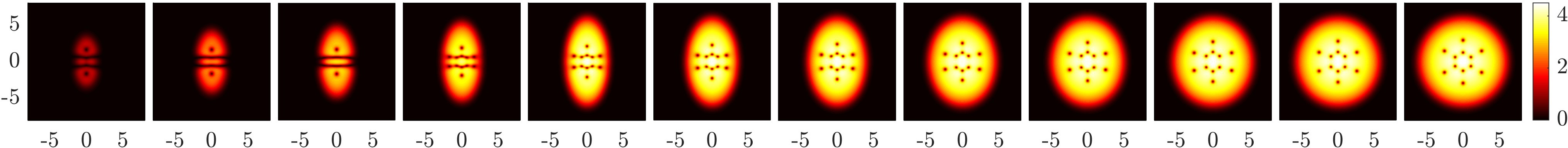}
\includegraphics[width=\textwidth]{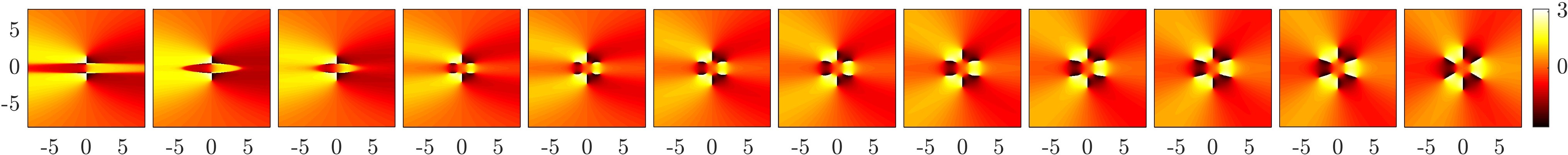}
\includegraphics[width=\textwidth]{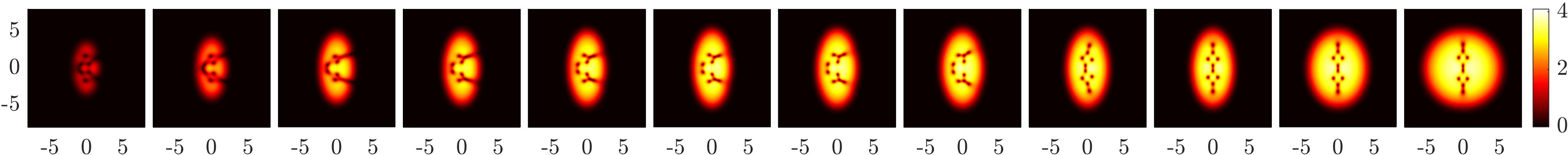}
\includegraphics[width=\textwidth]{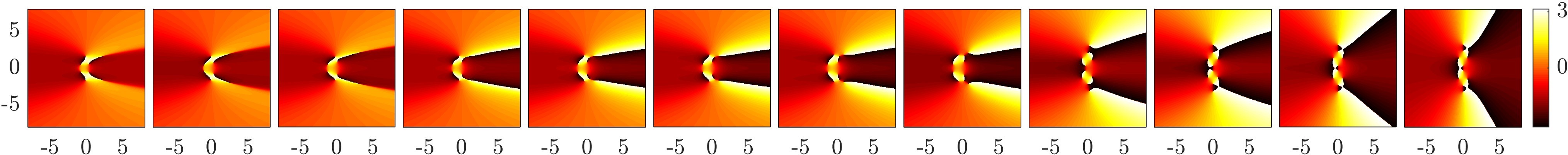}
\caption{
Solitary waves continued from the $\kappa=1/2$ linear degenerate set of $\mu_0=5.5$, continued from Fig.~\ref{DS04Aa}. For the VX6 state, typical states at $\mu=6.5, 11, 12, 16$ of $\omega_x=2$, and then $\omega_x=1.5, 1$ of $\mu=16$ are shown. Similarly, the states of VX8a at $\mu=6.5, 16$, and $\omega_x=1.5, 1.2, 1.1, 1$ are depicted. The second set illustrates the VX8b and VX8c states at $\mu=6.5, 10, 16$, and $\omega_x=1.7, 1.3, 1$. The third set shows the VX8d states at $\mu=6.5, 10, 13, 16, 20$, and $\omega_x=1.7, 1.5, 1.3, 1.2, 1.1, 1.05, 1$. The fourth set sketches the VX8e states at $\mu=6.5, 9, 12, 13, 14, 15$, and $\omega_x=1.9, 1.827, 1.826, 1.816, 1.3, 1$ of $\mu=15$. 
}
\label{VX6Aa}
\end{figure*}

The continuation of the DS12 state in both the chemical potential and the trap frequency is quite as expected. As the chemical potential increases, the two horizontal dark soliton stripes also connect at $\mu \approx 7.8$ in our spatial horizon. Therefore, the state evolves towards the $\phi$ soliton in the TF regime. As it is continued into the isotropic trap, it becomes the regular $\phi$ soliton of Fig.~\ref{DS30} as expected.

The DS20 state has a closed dark soliton loop in the phase profile directly in the near-linear regime, but the loop quickly opens around $\mu \lesssim 8$ and the two dark soliton stripes take approximately a hyperbolic shape. However, the dark soliton filaments become highly curved in the TF regime. The continuation in $\kappa$ is essentially clear, it becomes the regular DS20 state in the isotropic trap, but the detail is somewhat complicated as the equilibrium configuration depends nontrivially on $\kappa$. First, the filaments become less curved as $\omega_x$ is decreased, and then they close into a loop at about $\omega_x=1.914$, but it opens again quickly. Next, the dark soliton filaments connect into a closed loop, like the DS8 state below, at about $\omega_x=1.334$. We shall encounter this transition again for a few times, therefore, we name it as the $\Omega1334$ transition for convenience. Finally, the state gradually morphs into the regular DS20 state in the isotropic trap, cf. Fig.~\ref{DS20}.

The DS8 state has two neighbouring dark soliton loops facing each other, like the number $8$ or a sand glass. The continuation in the chemical potential is quite robust. As the state is continued into the isotropic trap, the two dark soliton loops reconnect into a single loop at about $\omega_x=1.89$, then it gradually morphs into the regular DS20 state in the isotropic trap as well.


The U2 state or the double U state has an outer U soliton and an inner U soliton, it is a parity symmetry-breaking configuration. As the chemical potential increases, the outer U soliton opens a smaller angle while the inner U soliton opens a larger angle, pushing the two filaments closer in the TF regime. As the trap frequency $w_x$ decreases, the two U solitons together open a progressively wider angle similar to that of the U and $\Psi$ solitons in Fig.~\ref{GSAa}. Therefore, they gradually morph into the DS20 state at about $\omega_x=1.68$. Interestingly, the critical transition frequencies are all very similar. The DS20 state then undergoes the above $\Omega1334$ transition, and then gradually morphs into the regular DS20 state in the isotropic trap.

The U2O state has two opposite U solitons surrounding a central deformed RDS oriented along the soft direction, and its continuation in the chemical potential is robust. These filaments become somewhat sharper and closer along the $y$ axis in the TF regime. As the trap frequency $w_x$ decreases, they undergo reconnections at about $\omega_x=1.79$, forming a DS20 isomer state. This highly deformed DS20 state becomes more regular and then also undergoes the $\Omega1334$ transition. Interestingly, the two DS20 states remain different but converge to essentially the same state before the transition. After the transition, the final state and the subsequent evolution become identical, yielding the DS20 state in the isotropic trap.

The VX6 state has $6$ vortices in a deformed ring shape in the low-density regime, $4$ inside the condensate and $2$ at the edge, as illustrated in Fig.~\ref{VX6Aa}. As the chemical potential increases, the edge vortices are induced into the condensate at about $\mu=10.5$. However, two pair creations quickly follow inside the condensate at $\mu \approx 11.1$ to heal the density depletion regions, forming two Y solitons and consequently a VX10 state. The continuation in $\kappa$ is quite robust, and the vortex configuration remains largely unchanged in the process.

The VX8a state has two approximately parallel aligned vortex quadruple states, and the vortex charge alternates between the neighbouring vortices, like a miniature lattice. The continuation in the chemical potential is very robust. As it is continued into the isotropic trap, the two aligned vortex quadruples gradually bend outward, forming interestingly two perpendicular vortex quadruples in the isotropic trap. The bending process is, however, not uniform. It approximately starts and then also accelerates from $\omega_x \lesssim 1.2$. The continuation requires a finer step size, e.g., $\delta \omega_x=0.0005$ as the equilibrium configuration strongly evolves therein. This state is typically spectrally unstable in the isotropic trap \cite{Middelkamp:VX}.


The vortex charge of the VX8b state alternates along the upper ring and the lower ring, and also between the two central vortices. The continuation in both the chemical potential and the trap frequency is remarkably robust. While the equilibrium configuration evolves during the continuation, the basic structure remains essentially the same.

The VX8c state has two Y vortex clusters facing each other, the vortex charge alternates between the neighbouring vortices of the two Y clusters. Similar to the VX8b state, the continuation in both the chemical potential and the trap frequency is particularly robust. This configuration also appears as the central part of the VX6 state above in the TF regime.

The VX8d state has three vortices in each of the density depletion regions and the vortex charge alternates between the neighbouring vortices therein, however, each of the two prominent vortices has the same charge as its central neighbour. As the chemical potential increases, each central vortex undergoes an elongation and pair creation at about $\mu=13.0$, leading to a total of $12$ vortices in the TF regime. The continuation in $\kappa$ is quite complicated, it only works until $\omega_x \approx 1.3943$ at $\mu=16$, reaching an existence boundary. However, the lower bound of $\omega_x$ decreases as the chemical potential increases. As such, we have continued this state to a slightly larger chemical potential $\mu=20$, which is sufficient for a successful continuation into the isotropic trap. The state in the isotropic trap has two concentric vortex necklaces of size $6$, like a benzene molecule. This state demonstrates that a solitary wave can have a nontrivial existence region in the $(\mu,\omega_x)$ parameter space.

The VX8e state has a parity symmetry-breaking structure like a deformed VX8a state in the small-density regime. As the chemical potential increases, the edge vortices are ejected out of the condensate at $\mu \approx 12.6$, leading to a temporary VX6 state. Two pair creations quickly follow at $\mu \approx 13.3$ to heal the density depletions, yielding a VX10 state in the TF regime. The continuation in $\kappa$ is rather challenging, the two Y clusters rotate oppositely towards the $y$ axis as $\omega_x$ decreases. It works until only $\omega_x \approx 1.8324$ at $\mu=16$, reaching again an existence boundary. In addition, this lower bound decreases rather marginally as the chemical potential decreases, showing the complexity of the trap continuation for some solitary waves. Fortunately, we have successfully captured a transition chaos from $\omega_x=1.827$ to $\omega_x=1.826$ at $\mu=15$. The two Y clusters are almost opposite to each other immediately after the transition, and the vortex dipole sits approximately at the center of the condensate. Then, the state quickly takes a more symmetric form for $\omega_x \lesssim 1.816$. The continuation thereafter is very robust, and we have also continued the final state in the isotropic trap to $\mu=16$ again. The $\mu=16$ states are very similar to the $\mu=15$ ones, therefore, they are omitted here for clarity. The final configuration is like a VX8c state with a central vortex dipole.

\begin{table}
\caption{
Number of distinct solitary waves $\Omega$ identified in the near-linear regime by the linear limit continuation method. Here, $g$ is the degeneracy of the linear degenerate set of $\kappa$ and $\mu_0$, $\Omega_r$ and $\Omega_c$ are the numbers of real-valued dark soliton states and the complex-valued states with vortices, respectively. The number of independent runs of the RRS and CRS is also summarized here for convenience in the last two columns, respectively.
\label{para}
}
\begin{tabular*}{\columnwidth}{@{\extracolsep{\fill}} l c c c c c c r}
\hline
\hline
$\kappa$ &$\mu_0$ &$g$ &$\Omega_r$ &$\Omega_c$ &$\Omega$ &RRS &CRS \\
\hline
$1$  &$1$ &$1$ &$1$ &$0$ &$1$ &- &- \\
\hline
$1$  &$2$ &$2$ &$1$ &$1$ &$2$ &$100$ &$100$ \\
\hline
$1$  &$3$ &$3$ &$3$ &$3$ &$6$ &$1000$ &$1000$ \\
\hline
$1$  &$4$ &$4$ &$4$ &$10$ &$14$ &$1000$ &$1000$ \\
\hline
$1$  &$5$ &$5$ &$9$ &$24$ &$33$ &$1000$ &$2000$ \\
\hline
$1/2$  &$1.5$ &$1$ &$1$ &$0$ &$1$ &- &- \\
\hline
$1/2$  &$2.5$ &$1$ &$1$ &$0$ &$1$ &- &- \\
\hline
$1/2$  &$3.5$ &$2$ &$3$ &$1$ &$4$ &$1000$ &$1000$ \\
\hline
$1/2$  &$4.5$ &$2$ &$3$ &$1$ &$4$ &$1000$ &$1000$ \\
\hline
$1/2$  &$5.5$ &$3$ &$6$ &$6$ &$12$ &$1000$ &$1000$ \\
\hline
\hline
\end{tabular*}
\end{table}

The continuation of the $\kappa=1/2$ linear degenerate sets is also a success. While some solitary waves are duplicated ones, numerous wave patterns are novel even when they are continued into the isotropic trap. Note that we have again limited our exploration to reasonably low-lying states, and more solitary waves are expected from the more excited linear degenerate sets. We find that $\kappa$ is a nontrivial parameter for the existence of some solitary waves. For example, both the U and $\Psi$ solitons do not survival in the continuation into the isotropic trap. The VX8a and VX8d states also look very different in the $\kappa=1/2$ and $\kappa=1$ traps, and the continuation in $\kappa$ of the VX8e state even suffers a discontinuous transition chaos. Future work should study the existence region in the $(\mu, \omega_x)$ parameter space for some typical solitary waves.

Finally, we summarize the number of solitary waves identified in the near-linear regime for the different linear degenerate sets investigated in Table~\ref{para}. The total number of solitary waves $\Omega$ grows very rapidly with increasing degeneracy $g$ for both potential traps. The exact scaling cannot be extracted accurately now, future work should study more excited states, but clearly it is strongly superlinear. The degeneracy grows slower for $\kappa=1/2$ because the neighbouring lattice points in a plane are further apart, however, more solitary waves appear to be available for the same degeneracy $g>1$. Furthermore, both the real-valued dark soliton states and the complex-valued states with vortices separately grow rapidly with increasing $g$. The number of independent runs of the RRS and CRS is also summarized here for convenience.





Our method is fundamentally different from the pioneering deflation method \cite{Panos:DC1,Panos:DC2,Panos:DC3}, the idea and the classification of the obtained states are both very different. The deflation method makes nontrial jumps from one state to another, which are not necessarily parametrically connected. As such, a set of states of mixed complexity is typically found \cite{Panos:DC1}. By contrast, all the bifurcations herein are genuinely from the linear limit. Therefore, our states stemming from a linear degenerate set have a similar complexity. Our method is also technically simpler, as different linear degenerate sets and the identified states therefrom can be studied independently with \textit{regular} computational methods. Clearly, the two methods can be developed in parallel, and also each of the ULSs herein can be applied as a starting point of the deflation method to accelerate the search of distinct solitary waves, their continuation and bifurcation diagrams.


\section{Summary and outlook}
\label{co}
In this work, we presented a coherent theoretical and numerical framework to systematically identify and continue solitary waves from their linear limits. First, we classify the possible linear degenerate states of a harmonic potential using lattice planes. Then, the distinct nonlinear waves bifurcating from a linear degenerate set are found using a random searcher and their underlying linear states are computed in the near-linear regime. These states are subsequently continued in the chemical potential into the TF regime. For states bifurcating from anisotropic traps, we also continue them into the isotropic trap in the TF regime. The method is applied to the two-dimensional scalar BECs in two prototypical traps $\kappa=1$ and $\kappa=1/2$, finding a very large number of solitary waves systematically. Their chemical potential evolution and the trap evolution for the latter setting are discussed in detail, including the typical continuation chaos processes. The number of solitary waves grows very rapidly with increasing degeneracy. The method is conceptually clear, computationally effective, and massively parallel. 

This work can be extended in numerous directions in the next. First, it is natural to continue more solitary waves from the other lattice planes. For example, the low-lying aspect ratios of $\kappa=1/3, 2/3, 1/4, 3/4, 1/5, 2/5, 3/5, 4/5$ are particularly interesting. Second, a systematic study of a few more excited linear degenerate sets is also of considerable interest. Considering the success of the method herein, we expect to find a lot more solitary waves in the future. Our preliminary results are in line with this expectation. While lots of solitary waves have been found in the literature, it seems likely this is just the beginning, and it is far away from the end even in the extensively studied $2d$ setting. A fun application of our work is that the solitary wave configurations can serve as wall papers, as more wave patterns are identified and continued. It is also interesting whether the box potential, which also has an analytic linear limit and it is experimentally relevant \cite{BoxP,Angel:VX4}, yields novel solitary wave structures. This is possible because wave patterns can be strongly influenced by the potential. A continuation in the rotating frequency is likely also relevant and interesting \cite{fetter2,RCG:rotating}.

It is also straightforward to apply the method to $3d$, at least conceptually, despite this is inevitably more computationally demanding. To our knowledge, the theoretical framework remains essentially the same. Here, one can again work with the Cartesian basis states of a $3d$ harmonic oscillator, and then the linear degenerate states can be classified using $3d$ lattice planes, focusing similarly on the $n_x, n_y, n_z \geq 0$ octant. As brute force calculations of $3d$ states are steadily emerging, this appears to be realistic in the near future, at least for low-lying states \cite{Ionut:VR,Wang:DSS,Panos:DC3,Brand_SV}.

It is also particularly interesting to extend the method to find vector solitary waves in multi-component BECs \cite{Panos:DC2,Wang:VRB}. In fact, the linear limit continuation method was initially developed for systematically constructing vector solitons in $1d$ in multiple earlier works \cite{Wang:DD,Wang:MDDD,Wang:DAD}. The techniques therein can be readily combined with the work here in $2d$ and $3d$. For example, a similar random solver can be designed for the two-component system, coupling two states from two distinct linear degenerate sets in the near-linear regime. Then, the vector state can be similarly continued into the TF regime in the vector chemical potential parameter space \cite{Wang:DD,Wang:DAD}.


Finally, the detailed properties of the solitary waves are worth studying in their own rights theoretically, computationally, and experimentally. Here, one can investigate their equilibrium configurations at the reduced particle level \cite{Wang:AI,PK:DSVX}, their experimental implementations, stability properties, and dynamics. When the solitary waves are unstable, it is interesting to study their typical decay scenarios, and explore whether they can be fully stabilized by tuning the chemical potential and the trap or adding suitable external potential barriers. Research efforts along some of these directions are currently in progress and will be reported in future publications.

\section*{Acknowledgments}
We gratefully acknowledge supports from the National Science Foundation of China under Grant No. 12004268, the Fundamental Research Funds for the Central Universities, China, and the Science Speciality Program of Sichuan University under Grant No. 2020SCUNL210.
We thank the Emei cluster at Sichuan
University for providing HPC resources.




\bibliography{Refs}

\end{document}